\documentclass[prd,twocolumn,
amsmath,
amssymb,
aps,
preprintnumbers,nofootinbib]{revtex4-1}
\pdfoutput=1
\usepackage{graphicx}% Include figure files
\usepackage{dcolumn}% Align table columns on decimal point
\usepackage{bm}% bold math
\usepackage{xcolor}
\usepackage[normalem]{ulem}
\usepackage{booktabs}
\usepackage{multirow}
\usepackage[export]{adjustbox}
\usepackage{floatrow}
\newfloatcommand{capbtabbox}{table}[][3.5in]
\newfloatcommand{capbfigbox}{figure}[][2.3in]
\usepackage{blindtext}
\usepackage{booktabs}
\usepackage{hepunits}
\usepackage[normalem]{ulem}
\usepackage{bbm}

\renewcommand{\GeV}{\,\text{GeV}}
\renewcommand{\MeV}{\,\text{MeV}}

\renewcommand{\vec}[1]{\mathbf{#1}} % vectors are boldface
\newcommand{\abs}[1]{\ensuremath{\left|#1\right|}}
\renewcommand{\eqref}[1]{Eq.~(\ref{#1})}
\DeclareMathOperator\sinc{sinc}
\newcommand{\BB}{\ensuremath{\mathbbm{B} }}
\newcommand{\MM}{\ensuremath{\mathbbm{M} }}

\begin{document}

\title{Dark Matter-Electron Scattering from Aromatic Organic Targets}

\author{Carlos Blanco$^{a,b}$}%\note{ORCID: http://orcid.org/0000-0001-8971-834X}
\email{carlosblanco2718@uchicago.edu}
\author{J.I. Collar$^{a,b}$}%\note{ORCID: http://orcid.org/XXXXXXX}
\email{collar@uchicago.edu}
\author{Yonatan Kahn$^{c}$}%\note{ORCID: http://orcid.org/XXXXXXX}
\email{yfkahn@illinois.edu}
\author{Benjamin Lillard$^{c}$}%\note{ORCID: http://orcid.org/XXXXXXX}
\email{blillard@illinois.edu}

\affiliation{$^a$University of Chicago, Department of Physics, Chicago, IL 60637}
\affiliation{$^b$University of Chicago, Kavli Institute for Cosmological Physics, Chicago, IL 60637}
\affiliation{$^c$University of Illinois at Urbana-Champaign, Department of Physics, Urbana, IL 61801}

\date{\today}% It is always \today, today,

% \begin{abstract}
%     Sub-GeV dark matter (DM) scattering with electrons can produce a detectable signal in a liquid scintillator. In particular, aromatic compounds such as benzene or xylene have an electronic excitation energy of a few eV, making them sensitive to DM as light as a few MeV. We develop the formalism for DM--electron scattering in aromatic organic molecules, calculate the expected rate in benzene and p-xylene, and apply this calculation to an existing measurement of single photo-electron emission rate in a low-background EJ-301 scintillator cell. Despite the fact that this measurement was performed in a shallow underground laboratory under minimal overburden, the DM--electron scattering limits extracted from these data are already approaching leading constraints in the 3--100 MeV DM mass range. We discuss possible next steps in the evolution of this direct detection technique, in which scalable organic scintillators are used in solid or liquid crystal phases and in conjunction with semiconductor photodetectors which would improve sensitivity through directional signal information and potentially lower dark rates per unit mass.
% \end{abstract}

\begin{abstract}
Sub-GeV dark matter (DM) which interacts with electrons can excite electrons occupying molecular orbitals in a scattering event. In particular, aromatic compounds such as benzene or xylene have an electronic excitation energy of a few eV, making them sensitive to DM as light as a few MeV. These compounds are often used as solvents in organic scintillators, where the de-excitation process leads to a photon which propagates until it is absorbed and re-emitted by a dilute fluor. The fluor photoemission is not absorbed by the bulk, but is instead detected by a photon detector such as a photomultiplier tube. We develop the formalism for DM--electron scattering in aromatic organic molecules, calculate the expected rate in p-xylene, and apply this calculation to an existing measurement of the single photo-electron emission rate in a low-background EJ-301 scintillator cell. Despite the fact that this measurement was performed in a shallow underground laboratory under minimal overburden, the DM--electron scattering limits extracted from these data are already approaching leading constraints in the 3--100 MeV DM mass range. We discuss possible next steps in the evolution of this direct detection technique, in which scalable organic scintillators are used in solid or liquid crystal phases and in conjunction with semiconductor photodetectors %which would 
to improve sensitivity through directional signal information and potentially lower dark rates.
\end{abstract}
\pacs{Valid PACS appear here}
\maketitle

\section{Introduction}

Dark matter (DM) can interact with electrons in a wide variety of systems, leading to a rich phenomenology of detector signatures and an active research program for development of new experiments  \cite{Essig:2011nj,Essig:2012yx,Graham:2012su,An:2014twa,Lee:2015qva,Essig:2015cda,Hochberg:2015pha,Hochberg:2015fth,Cavoto:2016lqo,Derenzo:2016fse,Hochberg:2016ntt,Emken:2017erx,Essig:2017kqs,Cavoto:2017otc,Fichet:2017bng,Tiffenberg:2017aac,Hochberg:2017wce,Romani:2017iwi,Agnes:2018oej,Crisler:2018gci,Agnese:2018col,Essig:2018tss,Abramoff:2019dfb,Aguilar-Arevalo:2019wdi,Aprile:2019xxb,Emken:2019tni,Coskuner:2019odd,Geilhufe:2019ndy,Trickle:2019nya,Griffin:2019mvc}. In particular, the formalism for DM--electron scattering in atoms \cite{Essig:2011nj} and solid-state systems \cite{Essig:2015cda,Trickle:2019nya} has been well-studied, but rather less attention has been devoted to DM--electron scattering in molecules.\footnote{See \cite{Arvanitaki:2017nhi}, however, for detailed studies of DM absorption in molecules, and \cite{Essig:2019kfe} for DM--nucleus scattering which excites rotational energy levels.} In principle, molecules are promising detector candidates because covalent molecular excitation energies can be comparable to semiconductor band gaps, $\mathcal{O}(\eV)$, allowing sensitivity to DM down to the MeV scale. Furthermore, achieving a large target mass (tens or hundreds of kg) of high-purity solvent such as benzene is somewhat easier than achieving a similar target mass of high-purity silicon, and when these molecules are used as solvents in a scintillating compound, the total background rate is quite competitive compared with the state-of-the-art in silicon achieved by SENSEI \cite{Tiffenberg:2017aac,Crisler:2018gci,Abramoff:2019dfb}, CDMS-HVeV \cite{Agnese:2018col}, and DAMIC at SNOLAB \cite{Aguilar-Arevalo:2019wdi}. Finally, because the scintillation signal is decoupled from the primary DM--electron scattering event, any photodetector sensitive to the wavelength of scintillation light may be used to read out the signal. This is in marked contrast with many DM--electron scattering proposals to date, including silicon CCD's, where the target material itself serves as the detector.

In this paper we set the first limits on DM--electron scattering from an organic scintillator target, specifically EJ-301 \cite{eljen}, a ternary scintillator composed of 95$\%$ p-xylene (1,4 dimethylbenzene) by mass, the rest being naphthalene and a proprietary fluor. We develop the theoretical formalism for DM--electron scattering in aromatic molecules (i.e. conjugated $\pi$-electron systems), exploiting the fact that semianalytic parametrizations of the electronic states can be found using a linear combination of atomic orbitals (LCAO) model. We then apply this formalism to the specific case of EJ-301, using experimental data from~\cite{Collar:2018ydf}, where a low-background 1.3 kg scintillator cell was operated in a dedicated shield under minimal overburden. The method described in \cite{Collar:2018ydf} allows
the dark count rate of the photomultiplier tube (PMT)
to be subtracted from the 
readout, resulting in a residual single photo-electron (SPE) rate of 3.8 Hz. As this background subtraction is only statistical in nature, and not event-by-event, it cannot be used to claim discovery, but nonetheless the residual rate is low enough to allow us to set an upper limit on the DM--electron cross section of 
about $10^{-34}$ cm$^2$ for 10 MeV DM scattering through a heavy mediator, within an order of magnitude of the world-leading limits set by DAMIC in this mass range. We note that our work is complementary to previous work on DM--electron scattering in scintillating targets \cite{Derenzo:2016fse}, which focused on solid-state systems as opposed to molecular solvents.

This paper is organized as follows. In Section~\ref{sec:Model}, we define our molecular model for aromatic compounds and determine the relevant electronic wavefunctions and excitation energies. In Section~\ref{sec:Rate}, we use these wavefunctions to compute the expected event rate for a given target mass. We present our results in Section~\ref{sec:Results}. In Section~\ref{sec:Future}, we discuss potential next steps in the development of organic scintillator targets, and we conclude in Section~\ref{sec:Conclusions} with some thoughts on how to scale this setup to achieve superior limits. Some technical details of the wavefunctions and molecular form factors can be found in Appendix \ref{app}.

\section{Molecular Orbital Model}
\label{sec:Model}
Dark matter can produce a detectable signal in a liquid scintillator by exciting an electron from the ground state into one of several low-lying unoccupied molecular orbitals, which fluoresce upon de-excitation. The emission lines are broadened by roto-vibrational energy sublevels, thermal motion, and solvent effects. Therefore, the emission spectra of a liquid scintillator will be a continuum with peaks corresponding to the electronic transitions to be discussed in this section. The emission spectrum for EJ-301 is presented in ref.~\cite{ej301}. 

A prediction for the scattering rate requires knowledge of the momentum space wavefunctions of the bound state electrons in the molecule. In this section we find analytic expressions for these wavefunctions using a linear combination of atomic orbitals (LCAO).
We then include configurational interactions (i.e. electron-electron repulsion) to obtain experimentally accurate descriptions of these electronic states. % in which two of the first singly excited states are degenerate and transform as $E_{1u}$ while the other two transform as $B_{1u}$ and $B_{2u}$~\cite{Katz1971,Pariser1953a,murrell1963theory}.
% In order to compute the scintillation rate from scattering of dark matter and the electrons in organic scintillators, it is necessary to derive the momentum space wavefunctions of the electronic states which can be excited to produce fluorescence upon de-excitation.
% EJ-301 is a ternary scintillator composed of 95$\%$ p-xylene (1,4 dimethylbenzene) by mass with the rest being composed of naphthalene and a proprietary fluor. Thus the primary target which initiates the scintillation process is p-xylene, a substituted benzene derivative. The chemical structures of benzene and p-xylene are shown in Fig.~\ref{fig:struct}.

\begin{figure}
    \centering
    \includegraphics[width=17.5mm]{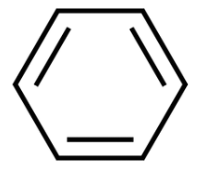}\;\;\;
    \includegraphics[width=30.5mm]{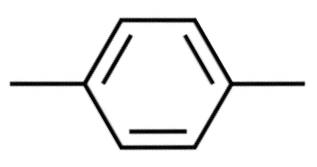}
    \caption{The chemical  structure of benzene (left) and p-xylene (right). Following the convention common in organic chemistry, vertices are taken to be carbon atoms, single lines are carbon--carbon single bonds, and double lines are carbon--carbon double bonds. The additional horizontal lines in p-xyelene represent two CH$_3$ methyl groups. In this case only one of the resonant Kekul\'{e} forms are shown for each molecule, but the rings should be understood to be entirely conjugated $\pi$-systems.}
    \label{fig:struct}
\end{figure}

In EJ-301, the primary target which initiates the scintillation process is p-xylene, a substituted benzene derivative with a chemical structure shown in Fig.~\ref{fig:struct} (right). In molecules such as benzene, the ring contains alternant double bonds. The electrons which occupy the $\pi$-orbitals in these bonds are essentially delocalized and are free to move along the so-called \textit{aromatic} ring made up of the conjugated $\pi$-bonds.\footnote{Single covalent bonds between carbon are formed when two electrons occupy the $\sigma$-bonding orbital between two $sp^3$-hybridized carbon atoms. The $\sigma$-orbital lies along the axis between the nuclei. In a carbon double bond, the carbons atoms are in an $sp^2$-hybridized state where %along with the $\sigma$-bonding orbital, the $\pi$-orbital is also occupied. 
both the $\sigma$-bonding orbital and the $\pi$-orbital are occupied.
The $\pi$-orbital is lobed above and below the molecular plane,
and is less tightly bound than the $\sigma$ orbital. 
}
In six-membered aromatic rings there are six electrons associated with the conjugated $\pi$-electron system.

%\footnote{
%The two electrons in a single covalent bond occupy a $\sigma$ orbital, which is aligned along the axis connecting the two nuclei. A double covalent bond includes an additional pair of electrons occupying a $\pi$ orbital, which is more easily excited or ionized than the tightly-bound $\sigma$ orbitals. 
%}

 The electronic transitions responsible for scintillation are those in which electrons in the conjugated $\pi$-bonds of the aromatic ring are excited into higher energy molecular orbitals. Thus, we will restrict our characterization to the $\pi$-conjugated system whose H$\text{\"u}$ckel molecular orbitals are linear combinations of six $2p_z$ atomic orbitals, one from each carbon in the ring. Each such carbon contributes one electron into the system. Following the well-known LCAO method~\cite{murrell1963theory}, the H$\text{\"u}$ckel molecular orbitals (HMO's), $\psi_i$, are given by
\begin{equation}
    \Psi_i = \sum_{j=1}^{6} c_{i}^{j} \phi_{2p_z}(\mathbf{r}-\mathbf{R}_j),
    \label{eq:LCAO}
\end{equation}
where $c_i$ are the coefficients to be determined and $\phi_{2p_z}(r-\mathbf{R}_i)$ are atomic orbitals of the Slater type and $\mathbf{R}_i$ are the equilibrium locations of the carbon nuclei. The $2p_z$ Slater atomic orbital is given by 
\begin{equation}
\label{eq:phidef}
    \phi_{2 p_{z}}(\mathbf{r})=\sqrt{\frac{Z^{3}_\text{eff}}{2^5   \pi a_{0}^{3}}}  \frac{ r \cos \theta}{a_{0}} \exp \left(\frac{-Z_{\text {eff }} r}{2 a_{0}}\right),
\end{equation}
where $a_0$ is the Bohr radius, and $Z_\text{eff} = 3.15$ is the effective nuclear charge of the carbon $2p_z$ orbital \cite{Pariser1953}.

The HMOs can be determined by diagonalizing the six-by-six Hamiltonian, which reduces to solving the following linear system,
\begin{equation}
    \sum_{l=1}^{6}\left[\left(H_{l m}-E_l \delta_{l m}\right) c_{i}^{l}\right]=0, \text { for } m=1,2, \ldots, 6
\end{equation}
where $H_{l m} = \langle \phi_l |\mathcal{H}_{core}| \phi_m \rangle$ are the Hamiltonian matrix elements. We follow a type of H$\text{\"u}$ckel model in which only nearest-neighbor interactions are considered. The Hamiltonian is then comprised of two types of matrix elements, one diagonal and one off-diagonal energy. The diagonal elements, i.e.~the onsite energy, is an empirical quantity which varies by elemental atom. The off-diagonal resonance integral is a measure of the nearest-neighbor nuclear interactions. The values for the coefficients are given in Appendix \ref{sec:coeff}.  %Here, the onsite energy is taken to be $\mathcal{E}_c=-6.7\text{eV}$ and nearest-neighbor resonance integral given by the following~\cite{Hawke2008, harrison2012electronic},
%\begin{equation}
%    H_{i j}=-0.63\frac{\hbar^{2}}{m_{e} d_{i j}^{2}}\delta_{i,i\pm 1} = -2.45\text{eV},
%\end{equation}
%where $d_{ij}$ is the distance between atoms. 

Following the notation of~\cite{murrell1963theory}, we label the six HMOs in order of increasing energy as $\Psi_{2}$, $\Psi_{1}$, $\Psi_{1^{\prime}}$, $\Psi_{-1}$, $\Psi_{-1^{\prime}}$, and $\Psi_{-2}$, where in the ground state of the molecule the first three are occupied and the last three are unoccupied (see Fig.~\ref{fig:struct}). The minimum electronic excitation energy is from the highest unoccupied molecular orbital (HOMO) to the lowest unoccupied molecular orbital (LUMO), analogous to the valence--conduction gap in a semiconductor. The coefficients for these H$\text{\"u}$ckel eigenstates are given in Appendix \ref{sec:coeff}. In this construction, ($\Psi_{1}$, $\Psi_{1\prime}$) and ($\Psi_{-1}$, $\Psi_{-1\prime}$) are degenerate pairs. However, since the the multi-electron wavefunction must be anti-symmetric, it is given by linear combinations of Slater determinants, i.e. the normalized and antisymmetrized products of six HMO's. The ground state for benzene is then given by:

\begin{equation}
   \psi_G = |\Psi_{2} \overline{\Psi}_{2}\Psi_{1} \overline{\Psi}_{1}\Psi_{1\prime} \overline{\Psi}_{1\prime}|,
\end{equation}
where $|\Psi_{1},...,\Psi_{n}|$ is the antisymmetrized product of the HMOs ${\Psi_{1},...,\Psi_{n}}$, and $\overline{\Psi}$ is the opposite spin state as $\Psi$. In these products, the order specifies the identical electron indexing which implies $|\Psi_0 \overline{\Psi}_0| = |\Psi_0 (1) \overline{\Psi}_0 (2)| = -|\overline{\Psi}_0 (1)\Psi_0 (2)|$.

Following the method of Pariser, Pople, and Parr ~\cite{Pariser1953,Pariser1953a,pople1955electronic,pople1970molecular}, we now include the electron repulsion term in the Hamiltonian, which becomes:
\begin{equation}
    \mathcal{H}_{ppp} = \mathcal{H}_{core} + \sum_{ij}{\frac{e^2}{r_{ij}}},
\end{equation}
where $\mathcal{H}_{core}$ is the core Hamiltonian which gave us the HMOs and the second term is the two-electron repulsion operator responsible for the configurational energy of the electrons. In order to obtain the energies of each transition, the electron repulsion integrals must be calculated using the many-body wavefunctions~\cite{Pariser1953, Roothaan1951}. Here, we adopt a semi-empirical model in which the first singlet excitation energies are tuned to the experimental values for p-xylene~\cite{Katz1971}, and the second and third singlet excitation energies are calculated from the parameters derived in the literature~\cite{Pariser1953}. Note that since the sensitivity reach of the scintillator is dominated by the lowest-lying excitations it is not particularly sensitive to large uncertainties in $\Delta E^{s_2}$ and $\Delta E^{s_3}$. The molecular orbitals for the first singly excited singlet states are given by the following \cite{Wagniere1976,murrell1963theory},
\begin{equation}
\begin{aligned}
    \psi_{1}^{s_1} &= 1/\sqrt{2}\; (\psi_{-1}^{1\prime}-\psi_{-1\prime}^{1}),\; \Delta E_{1}^{s_1} = 4.5~\eV\\
    \psi_{2}^{s_1} &= 1/\sqrt{2}\; (\psi_{-1\prime}^{1\prime}+\psi_{-1}^{1}),\; \Delta E_{2}^{s_1} = 5.6~\eV\\
    \psi_{3}^{s_1} &= 1/\sqrt{2}\; (\psi_{-1}^{1\prime}+\psi_{-1\prime}^{1}),\; \Delta E_{3}^{s_1} = 6.4~\eV\\
    \psi_{4}^{s_1} &= 1/\sqrt{2}\; (\psi_{-1\prime}^{1\prime}-\psi_{-1}^{1}),\; \Delta E_{3}^{s_1} = 6.4~\eV,
\end{aligned}
\label{eq:firstexcited}
\end{equation}
where 
\begin{equation}
    \psi_i ^j = \frac{1}{\sqrt{2}} (|\Psi_{1} \overline{\Psi}_{1}...\Psi_{i} \overline{\Psi}_{j}...\Psi_{N} \overline{\Psi}_{N}| - |\Psi_{1} \overline{\Psi}_{1}...\Psi_{j} \overline{\Psi}_{i}...\Psi_{N} \overline{\Psi}_{N}|),
\end{equation}
are single electron singlet excitations with respect to the ground state. The second and third singly excited singlet excitation wavefunctions and energies are given in Appendix \ref{sec:excit}.
Following the notation of~\cite{Katz1971,Pariser1953a,murrell1963theory}, the $\psi_1^{s_1}$ and $\psi_2^{s_1}$ orbitals transform in the $B_{1u}$ and $B_{2u}$ representations of the point symmetry group, while the degenerate $\psi_3^{s_1}$ and $\psi_4^{s_1}$ transform as $E_{1u}$.

The rate of dark matter--electron scattering depends on the molecular form factor,
\begin{align}
f_{i  j} (\vec{q})  &= \int d^3 \vec{p}\,  \tilde{\psi}_i (\vec{p}) \tilde{\psi}_j^\star (\vec{p}+\vec{q} )\\
%                    &= \langle \tilde{\psi}_j(\vec{p}+\vec{q}) | \tilde{\psi}_i (\vec{p})\rangle\\
                    &= \langle \psi_j(\vec{r}) |e^{i\vec{q} \cdot \vec{r}}|\psi_i (\vec{r})\rangle,
\label{eq:molformfactor}
\end{align}
which specifies the probability of transferring  momentum $\vec{q}$ to the molecule while exciting an electron from the initial state $\psi_i$ to the final state $\psi_f$ (here $\tilde{\psi}(\vec{k})$ are the momentum space wavefunctions). Since the molecular form factor is an inner product of single-electron operators, the molecular form factors between the ground state and the excited states is evaluated % using the following identity,
via
\begin{equation}
    \langle \psi_i^j(\vec{r}) |e^{i\vec{q} \cdot \vec{r}}|\psi_G (\vec{r})\rangle = \sqrt{2}\langle \Psi_j(\vec{r}) |e^{i\vec{q} \cdot \vec{r}}|\Psi_i (\vec{r})\rangle,
\end{equation}
allowing form factors over many-body wavefunctions, $\psi$, to be computed in terms of the single electron HMO's, $\Psi$.

\begin{figure}
    \centering
    \includegraphics[width=42mm]{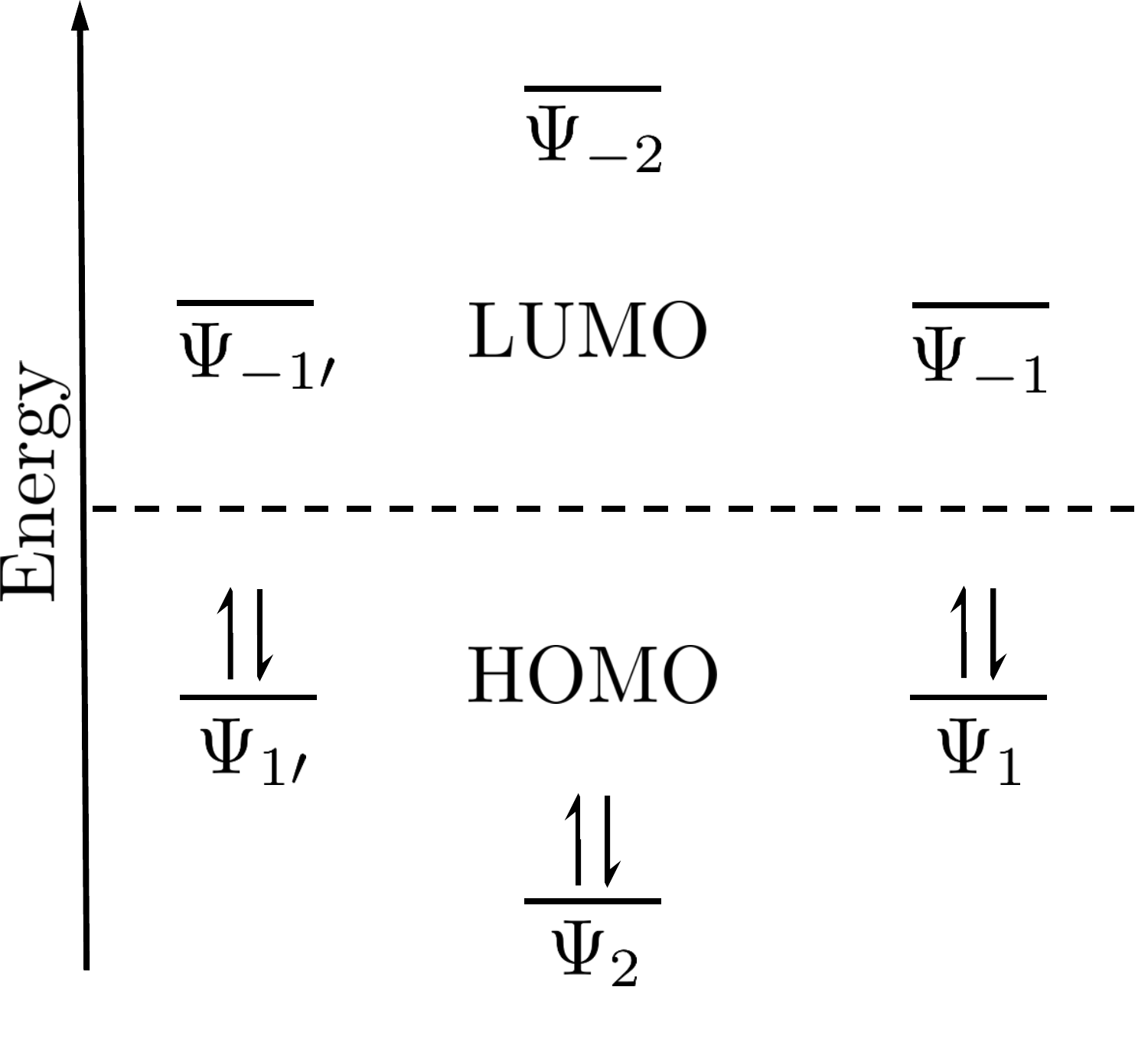}
    \includegraphics[width=42mm]{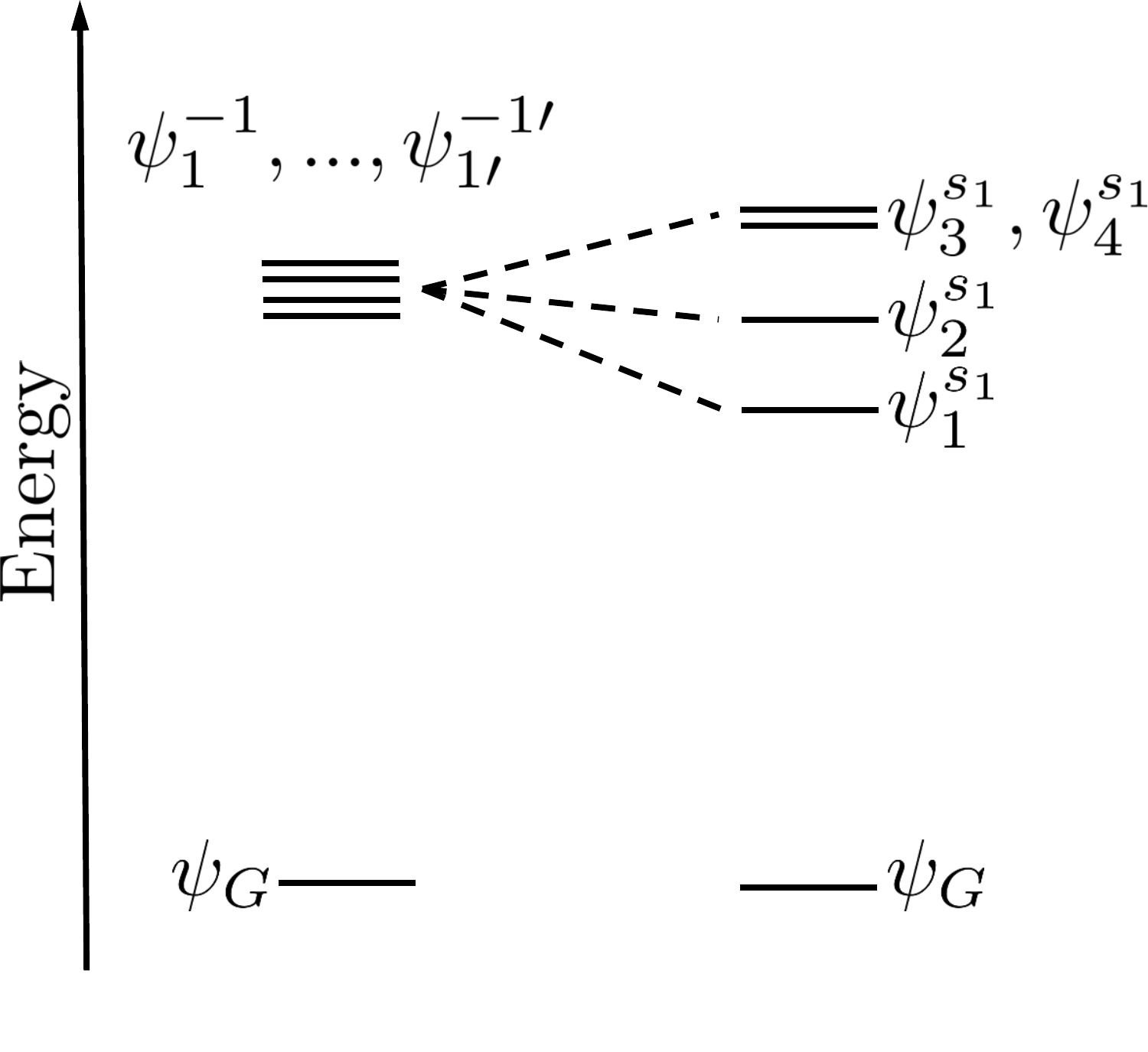}
    \caption{(Left) A schematic diagram of the energies of the HMO's of benzene following the notation of~\cite{murrell1963theory}. Six electrons occupy the three lowest orbitals up to the highest occupied molecular orbitals (HOMO). The lowest energy transitions are from a HOMO state to one of the lowest unoccupied molecular orbitals (LUMO). (Right) A schematic of the energy splitting brought about by the configurational interaction between electrons in the antisymmetrized MO's.}
    \label{fig:homolumo}
\end{figure}

% \begin{figure}
%     \centering
%     \includegraphics[width=15mm, angle =90]{benzene.png}\;\;\;
%     \includegraphics[width=15mm, angle =90]{p-xylene.png}
%     \caption{The chemical  structure of benzene (left) and p-xylene (right). Following the convention common in organic chemistry, vertices are taken to be carbon atoms, single lines are carbon-carbon bonds, and double lines are carbon-carbon double bonds. In this case only one of the resonant Kekul$\text{\'e}$ forms are shown for each, but the rings should be understood to be entirely conjugated $\pi$-systems}
%     \label{fig:struct}
% \end{figure}

Since the methyl substituents in p-xylene are $\sigma$-bonded, they do not affect the $\pi$-electron system to first order in our model, and so the predicted wavefunctions for the conjugated $\pi$-electrons in p-xylene are equivalent to those derived for benzene. We account for the presence of p-xylene in the scintillator by tuning the excitation energies to those listed in \eqref{eq:firstexcited}. Note that the procedure described in this section for obtaining the many-body wavefunctions of electrons in benzene can be used to describe any molecule with a conjugated $\pi$-electron system, including the aromatic compounds used in other types of scintillators.

\section{Rate Calculation}
\label{sec:Rate}

Dark matter with sufficient kinetic energy can induce a transition from the ground state to one of the excited bound states of~\eqref{eq:firstexcited} (or Appendix~\ref{sec:excit})
%~\eqref{eq:secondexcited} and~\eqref{eq:thirdexcited}
with a scattering rate determined in part by the momentum space wavefunctions of the initial and final electronic states.  
Conveniently, the details of molecular physics factorize into the form factor $f_{ij}(\vec{q})$ of \eqref{eq:molformfactor}, so that the scattering cross section 
$(\sigma v)_{i\rightarrow j}$ for any of the possible transitions reduces to \cite{Essig:2015cda}
\begin{align}
(\sigma v)_{i\rightarrow j} 
= \int \frac{d^3 \vec{q} }{(2\pi)^3 } (2\pi)\delta(\Delta E_{ij} - \omega ) \frac{\abs{ \mathcal M_\text{free} }^2 \abs{f_{ij}(\vec{q}) }^2 }{16 m_\chi^2 m_e^2},
\end{align}
where $\mathcal M_\text{free}$ is the amplitude for dark matter scattering on free electrons; $\vec{q}$ is the momentum transferred from the dark matter; $m_\chi$ and $m_e$ are the dark matter and electron masses; $i$ and $j$ label the initial and final states, with a difference in energy of $\Delta E_{ij}$; and
\begin{equation}
    \omega = \frac{q^2}{2m_\chi} - \mathbf{q} \cdot \mathbf{v}
\end{equation}
is the energy transferred from the DM to the p-xylene molecule for an incoming DM velocity $\mathbf{v}$ (we have approximated the reduced mass of the DM--xylene system as just $m_\chi$ which is appropriate for sub-GeV DM).
% In terms of the momentum space wavefunctions $\Psi = \tilde{\psi}(\vec{k}) \left| \vec{k} \right\rangle$, the molecular form factor is simply
% \begin{align}
% f_{i  j} (\vec{q}) = \int d^3 \vec{p}\,  \tilde{\psi}_i (\vec{p}) \tilde{\psi}_j^\star (\vec{p}+\vec{q} ).
% \label{eq:molformfactor}
% \end{align}
It is conventional to separate the momentum dependence of $\mathcal M_\text{free}$ into a model-dependent form factor $F_\text{DM}(q)$,
\begin{equation}
\frac{\abs{\mathcal M_\text{free}}^2 }{16 m_\chi^2 m_e^2 } \equiv \frac{\pi \bar\sigma_e}{\mu_{\chi e}^2} F_\text{DM}^2(q)
\end{equation}
written in terms of the reduced mass of the DM--electron system, $\mu_{\chi e}$, and the cross section $\bar \sigma_e$ evaluated at a specific value of the momentum conventionally taken to be $q = \alpha m_e$ where $\alpha$ is the fine-structure constant. $F_\text{DM} = 1$ corresponds to a heavy particle mediating the DM--electron interaction, while $F_\text{DM}(q) = (\alpha m_e)^2/q^2$ corresponds to a light mediator.
% The form factors approach unity at $F_\text{DM}(q = \alpha m_e)$ and $\abs{f_{i,i}(\vec{q} \rightarrow 0)}$.

A shift in the coordinate of a function $\vec{r} \rightarrow \vec{r} - \vec{R}$, as in \eqref{eq:LCAO}, simply adds a phase $e^{-i \vec{k}\cdot \vec{R}}$ to its Fourier transform.
This property makes the LCAO model an especially powerful tool for obtaining approximate analytic expressions for the molecular form factor. In the case of benzene or xylene, the linear combination of atomic orbitals produces 
\begin{align}
\widetilde{\Psi}(\mathbf{k}) =\left(\sum_{i=1}^{6} c^{\psi}_{i} e^{-i \mathbf{k} \cdot \mathbf{R}_{i}}\right) \widetilde{\phi}(\mathbf{k}) \equiv \mathcal{B}_{\psi}(\mathbf{k}) \widetilde{\phi}(\mathbf{k}),
\label{eq:Psitilde}
\end{align}
%\begin{equation}
%\begin{aligned} 
%&\widetilde{\Psi}(\mathbf{k}) =\sum_{i=1} c_{i}(2 \pi)^{-3 / 2} \int d^{3} \mathbf{r} e^{-i \mathbf{k} \cdot \mathbf{r}} \phi\left(\mathbf{r}-\mathbf{R}_{i}\right) \\ 
%%&=\sum_{i=1}^{6} c_{i}(2 \pi)^{-3 / 2} e^{-i \mathbf{k} \cdot \mathbf{R}_{i}} \int d^{3} \mathbf{r}_{i}^{\prime} e^{-i \mathbf{k} \cdot \mathbf{r}_{i}^{\prime}} \phi\left(\mathbf{r}_{i}^{\prime}\right) \\ 
%&=\left(\sum_{i=1}^{6} c_{i} e^{-i \mathbf{k} \cdot \mathbf{R}_{i}}\right) \widetilde{\phi}(\mathbf{k}) \equiv \mathcal{B}_{\psi}(\mathbf{k}) \widetilde{\phi}(\mathbf{k}),
%\end{aligned}
%\end{equation}
where $\tilde\phi(\vec{k})$ is the Fourier transform of the $2 p_z$ Slater type atomic orbital \eqref{eq:phidef}, 
\begin{equation}
\tilde \phi_{2 p_z}(\vec{k}) =  a_0^{3/2} \frac{\sqrt{2}}{\pi} Z_\text{eff}^{7/2}
 \frac{a_0\, k_z}{ \left( a_0^2 k^2 + (Z_\text{eff} /2)^2 \right)^3 },
\end{equation}
and where $\mathcal B_{\psi}$ is a  prefactor defined in terms of the coefficients $c_{i}^\psi$ for each of the six HMO's $\Psi_i$, the expressions for which are given in Appendix \ref{sec:coeff}~\cite{tamagawa1976molecular,khan2018fluorescence}.

The total scattering rate $R$ expected in the detector
% involves the sum over all nine distinct transitions from the occupied states to lowest or next-to-lowest unoccupied states, and also 
depends on details of the dark matter velocity distribution, $g_\chi (\vec{v})$:
\begin{align}
R = \sum_{i, j} N_B \frac{\rho_\chi}{m_\chi} \int d^3 \vec{v} g_\chi(\vec{v}) (\sigma v)_{ij},
\end{align}
where $\rho_\chi$ is the local dark matter density, and $N_B$ is the number of p-xylene molecules in the scintillator.
We approximate $g_\chi(\vec{v})$ as a spherically symmetric speed distribution, and upon defining 
\begin{align}
\eta(v^{ij}_\text{min}) &= \int \frac{4 \pi v^2 dv }{v} g_\chi(v) \Theta(v - v^{ij}_\text{min}(q)),
\\
v^{ij}_\text{min}(q) &= \frac{\Delta E_{ij} }{q} + \frac{q}{2 m_\chi},
\end{align}
the expected rate of scintillation photons in the detector is
\begin{align}
R = \xi \frac{ N_B \rho_\chi \bar \sigma_e}{8\pi m_\chi  \mu_{\chi e}^2} \sum_{i,j} \int \frac{d^3 \vec{q} }{q} \eta(v^{ij}_\text{min}(q)) F_\text{DM}^2(q)  \abs{ f_{i  j}(\vec{q}) }^2,
\label{eq:rate}
\end{align}
where we have inserted an efficiency factor $\xi$, which is the product of the radiative quantum yield of the scintillator (that is, the probability of a primary excitation yielding a scintillation photon which escapes the liquid) and the quantum efficiency of the photodetector. A measurement of $R$ puts an upper bound on $\bar\sigma_e$, after assuming a particular form for $F_\text{DM}(q)$ and integrating \eqref{eq:rate}. Note that this rate does \emph{not} include a contribution from electron ionization (as opposed to excitation to a bound state); while including ionization would only increase our rate estimates, the dynamics of free electrons in liquid scintillators and the corresponding scintillation rate estimate are much more involved than our simple treatment here, so to be conservative we neglect this contribution entirely.
% A measurement of $R$ puts a maximum bound on $\bar\sigma_e$, after assuming some form for $F_\text{DM}(q)$.

Taking advantage of the fact that the six carbon atoms in the benzene and xylene rings are coplanar, 
every $\mathcal B_\psi(\vec{k})$ can be written as a function only of some $k_x$ and $k_y$, allowing the $p_z$ integral in \eqref{eq:molformfactor} to be evaluated analytically.
%For the benzene ring the six $\vec{R}_i$ are coplanar, allowing every $\mathcal B_\psi(\vec{k})$ to be written as a function only of $k_x$ and $k_y$. The bond length is also uniform, so that $\abs{\vec{R}_i} = 0.14\nm$ for both benzene and p-xylene~\cite{tamagawa1976molecular,khan2018fluorescence}.
We perform the remaining integrals in \eqref{eq:molformfactor} and \eqref{eq:rate} numerically using a Python implementation of the VEGAS adaptive Monte Carlo algorithm~\cite{Lapage:1978}. Upon expanding $\abs{f_{ij}(\mathbf{q})}^2$ as the product of two independent integrands, the rate $R$ from \eqref{eq:rate} can be calculated relatively quickly from the resulting seven-dimensional integral.\footnote{In Appendix~\ref{appx:analytic}, we discuss how the molecular form factor can be evaluated as an infinite series involving a generalization of the hypergeometric function.}

\begin{figure}
\centering
\includegraphics[clip, trim=0cm 1cm 0cm 0cm,width=86mm]{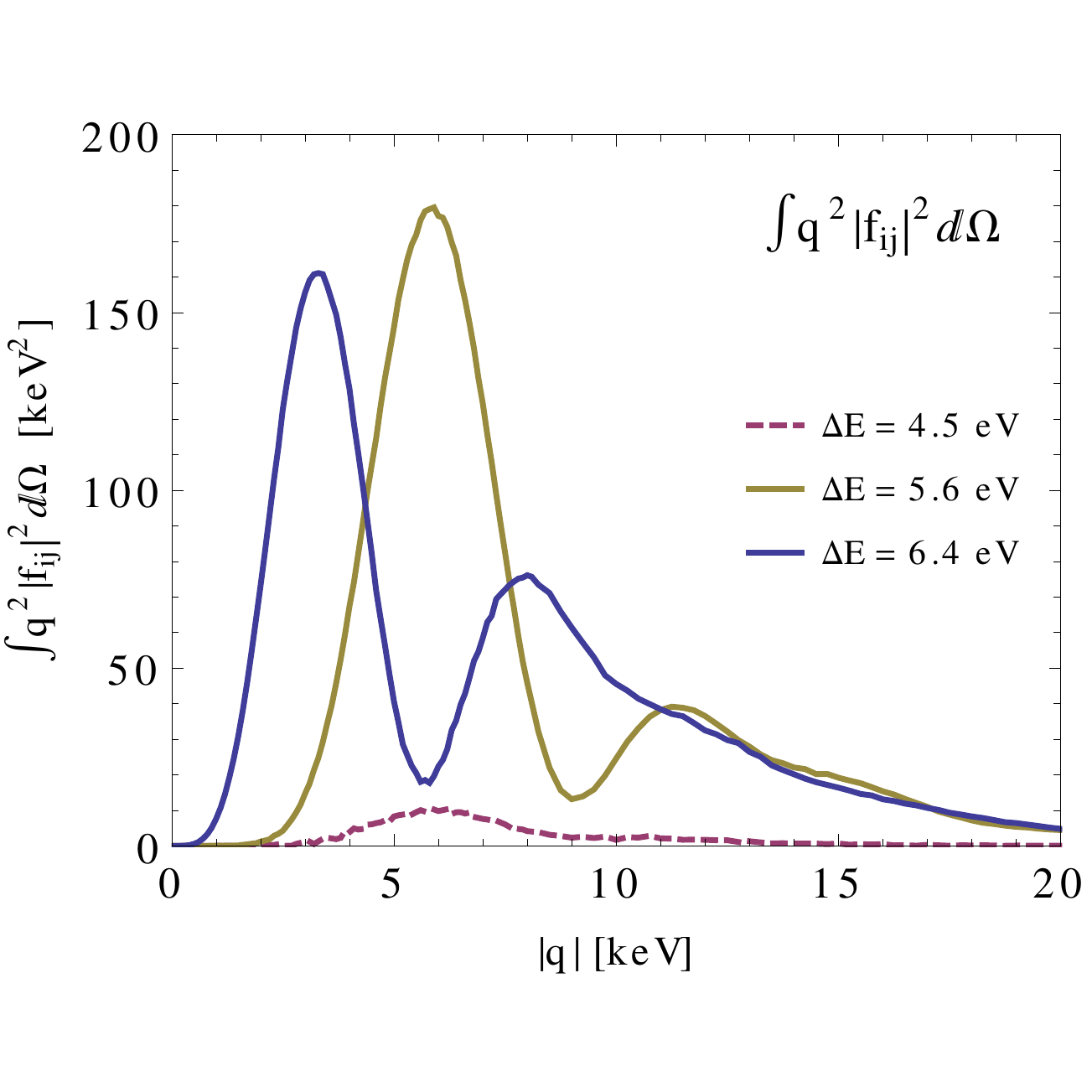}
\caption{Form factors squared $q^2 |f_{ij}(\vec{q})|^2$ for benzene integrated over angular variables, as a function of the magnitude of momentum transfer $|\vec{q}| = q$ for the three dominant transitions. Note that the lowest-energy transition at 4.5 eV (dashed) is strongly suppressed after angular integration.
}
\label{fig:molecularformfactor}
\end{figure}

In Fig.~\ref{fig:molecularformfactor}, we plot the angular integral of the absolute squares of the molecular form factors for each of the nine electronic transitions, i.e. $\int d\Omega_q q^2 |f_{ij}(\vec{q})|^2$. We note two important features: (a) the form factors peak at a momentum transfer, $q \sim 1/a_0 \approx 2.5~\keV$, as expected for atomic systems, and (b) perhaps surprisingly, the lowest-energy 4.5 eV transition is strongly suppressed, such that the effective gap is closer to 5.6 eV. As we show in Appendix \ref{sec:ffdetails}, the form factors exhibit strong directionality, which is washed out under the assumption of a spherically-symmetric DM distribution but would be relevant for the true DM velocity distribution in the Earth frame, for which the DM ``wind'' exhibits diurnal and annual modulation. This suggests the possibility, which we discuss further in Sec.~\ref{sec:Future}, of using benezene as a directional DM detector: indeed, organic molecules including one or more benzene rings can exhibit a liquid crystal phase, and by applying suitable electric fields, nematic liquid crystals can have the constituent molecules aligned along a particular direction over large distances. We leave a dedicated analysis of the possibilities of such a liquid crystal scintillator for directional DM detection to future work.

\section{Results}
\label{sec:Results}

In a previously reported measurement \cite{Collar:2018ydf}, a 1.5 liter (1.3 kg) cell of EJ-301 scintillator, specially developed to minimize internal radioactive backgrounds, was operated in a dedicated shield under a modest %(6.25 m.w.e.)
overburden of 6.25 meters water-equivalent (m.w.e). By reducing the temperature of the PMT photocathode it was possible to subtract the contribution from the PMT dark current to the SPE rate, generating in the process new limits on proton scattering  by few-GeV DM candidates \cite{Collar:2018ydf,PhysRevD.100.063013}. An irreducible SPE rate of $3.8 \pm 0.1$ Hz was obtained. Note that since this measurement was achieved over 14 days of data taking, comprising $4.6 \times 10^{6}$ events, the Poisson uncertainty of $\sim 2100$ events $= 0.0018$ Hz is far below the 0.1 Hz uncertainty coming from other sources such as temperature control. Thus, we set limits by simply scaling the 3.8 Hz rate. This setup was recently upgraded to include an improved temperature control and discrimination against delayed PMT afterpulses \cite{Collar:2018ydf}. The SPE background nevertheless remained unaltered, strongly suggesting that the present reach of the method is already limited by the environmental backgrounds associated to shallow underground operation.

To facilitate comparison with recent DM--electron scattering results, we take $\rho_\chi = 0.3 \ \GeV/{\rm cm}^3$ and evaluate the expected rate using \eqref{eq:rate} taking the speed distribution to be Maxwellian with a sharp cutoff at the escape velocity of the galaxy, 
\begin{equation}
g_\chi(v) = \frac{1}{K} \exp\left(- \frac{ \abs{ \vec{v} + \vec{v_E} }^2 }{v_0^2} \right) \Theta(v_\text{esc} - \abs{\vec{v} + \vec{v_E} } )
\end{equation}
where we take the typical velocity of the Earth to be $v_E = 232\text{ km/s}$, local mean speed $v_0 = 220\text{ km/s}$, and escape velocity $v_\text{esc} = 544\text{ km/s}$ \cite{Lewin:1995rx}.\footnote{We note that more sophisticated models of the halo exist which take into account updated measurements of the escape velocity and recent stellar data showing a radial anisotropy \cite{Evans:2018bqy}, but to perform an unbiased comparison we make the same choices as SENSEI, CDMS-HVeV, DAMIC, and XENON.} The quantum efficiency of the PMT  photocathode \cite{pmt}, integrated over the EJ-301 emission wavelengths \cite{ej301} is 21.5\%, and we calculate the radiative quantum yield of the scintillator to be 77\% (see Appendix \ref{sec:eff}), which gives $\xi \approx 0.166$. As discussed in \cite{Collar:2018ydf}, the light collection efficiency of the EJ-301 cell is compatible with 100\%.

\begin{figure}[t]
    \centering
    \includegraphics[width=88mm]{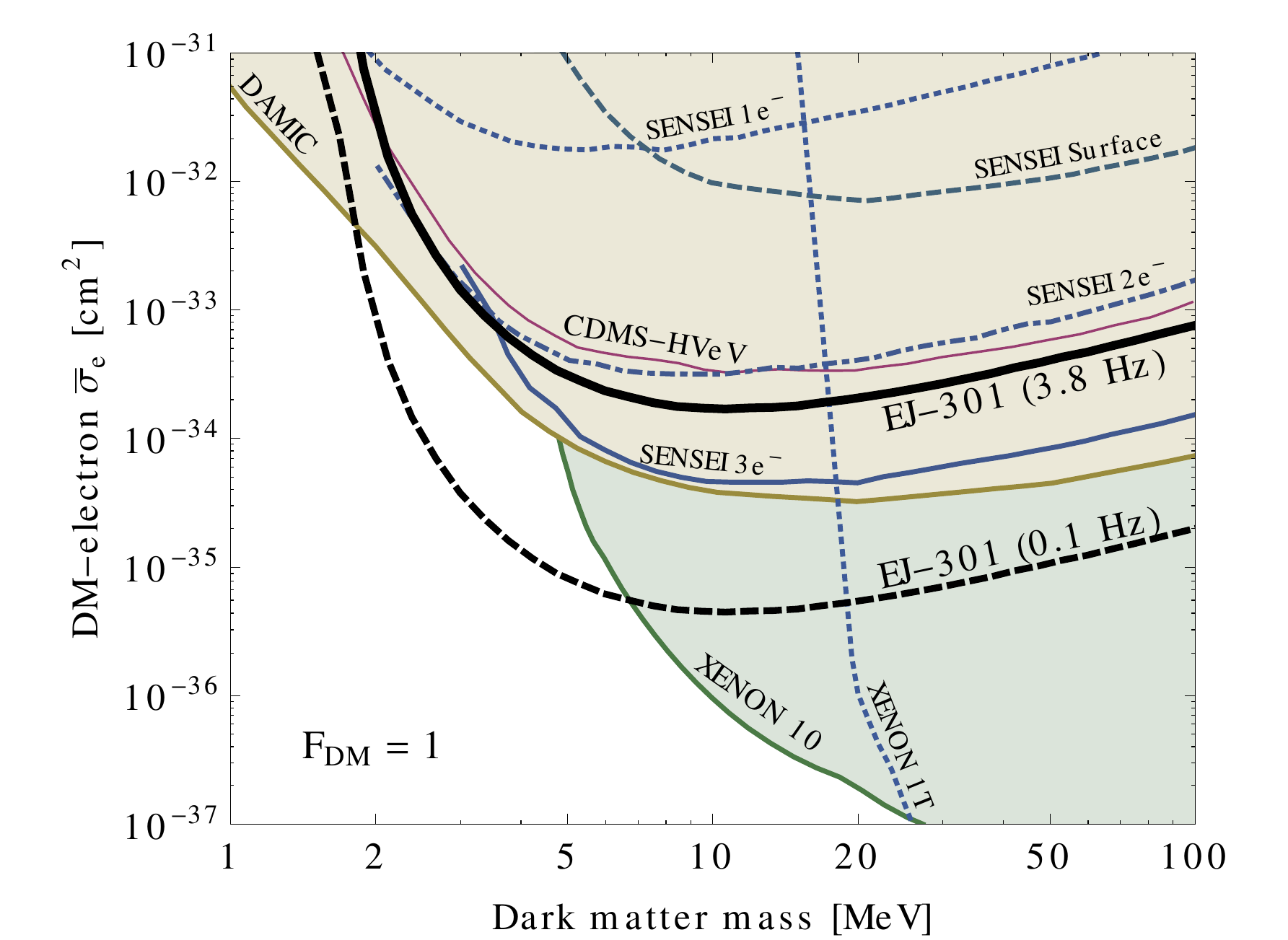}
    \includegraphics[width=88mm]{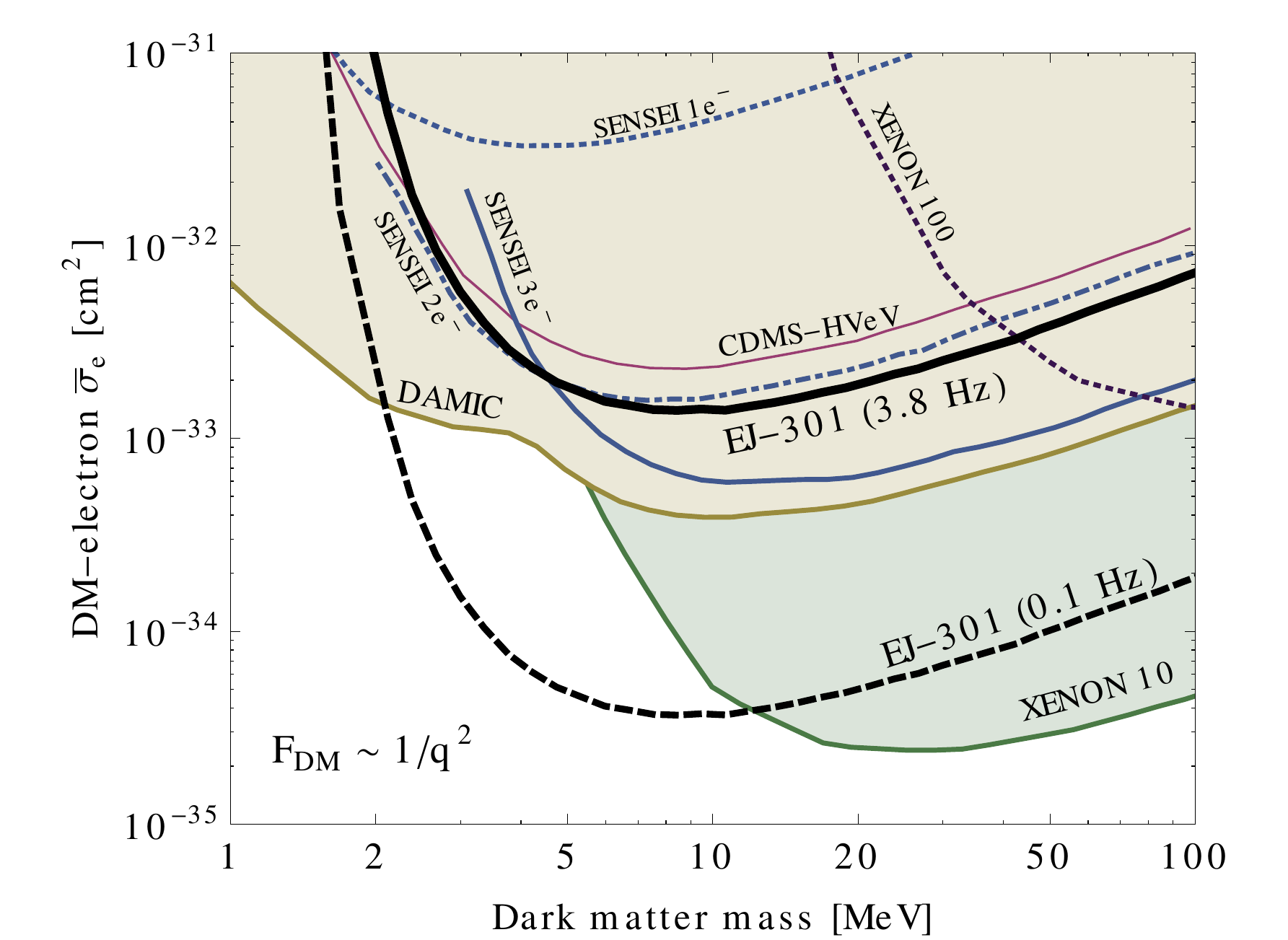}
    \caption{Limits on sub-GeV DM--electron scattering, for dark matter form factors $F_{\rm DM}=1$ (top) and $F_{\rm DM} \sim 1/q^2$ (bottom). The solid black curve shows the limit derived from the 3.8 Hz residual background of 1.3 kg of EJ-301 scintillator operated in a shallow (6.25 m.w.e.) laboratory \cite{Collar:2018ydf}, and the dashed black curve shows the potential improvements from a background rate of 0.1 Hz. Shaded regions show existing exclusions from DAMIC~\cite{Aguilar-Arevalo2019}, SENSEI~\cite{Tiffenberg:2017aac,Crisler:2018gci,Abramoff:2019dfb}, CDMS-HVeV \cite{Agnese:2018col}, XENON10/100 \cite{Essig:2012yx}, and XENON1T \cite{Aprile:2019xxb}, rescaled as needed to a common DM density of $\rho_\chi = 0.3~\GeV/\text{cm}^3$. In the bottom panel we show the stronger limit for XENON10 following the analysis of~\cite{Essig:2017kqs}. }
    \label{fig:exclusion}
\end{figure}

Fig.~\ref{fig:exclusion} shows the DM--electron scattering limit from EJ-301, conservatively assuming that 100\% of the irreducible 3.8 Hz rate is due to DM interactions. Factoring out the 21.5\% quantum efficiency of the PMT, the input-referred background rate is 17.7 Hz. Despite the large background, the limit for both $F_{\rm DM} = 1$ and $F_{\rm DM}=\alpha^2 m_e^2/q^2$ is quite competitive with the recent DAMIC \cite{Aguilar-Arevalo2019} and SENSEI \cite{Abramoff:2019dfb} constraints, even exceeding the limits from XENON 10 \cite{Essig:2017kqs} and XENON1T \cite{Aprile:2019xxb} below 5 MeV due to the lower threshold (5.6 eV for the lowest unsuppressed excited state in EJ-301 versus the 13 eV ionization energy of Xe) and surpassing the prototype CDMS-HVeV run \cite{Agnese:2018col} for $m_\chi > 3 \ \MeV$. A modest improvement in the background rate, which could potentially be achieved either with additional overburden to reduce the cosmic rate or by using a photodetector with a lower dark rate, could set world-leading limits on DM--electron scattering in the mass range 2 -- 7 MeV. We illustrate this in Fig.~\ref{fig:exclusion} by assuming a hypothetical background rate reduction to 0.1 Hz, giving the dashed black curve.

\begin{figure}[t]
    \centering
    \includegraphics[width=88mm]{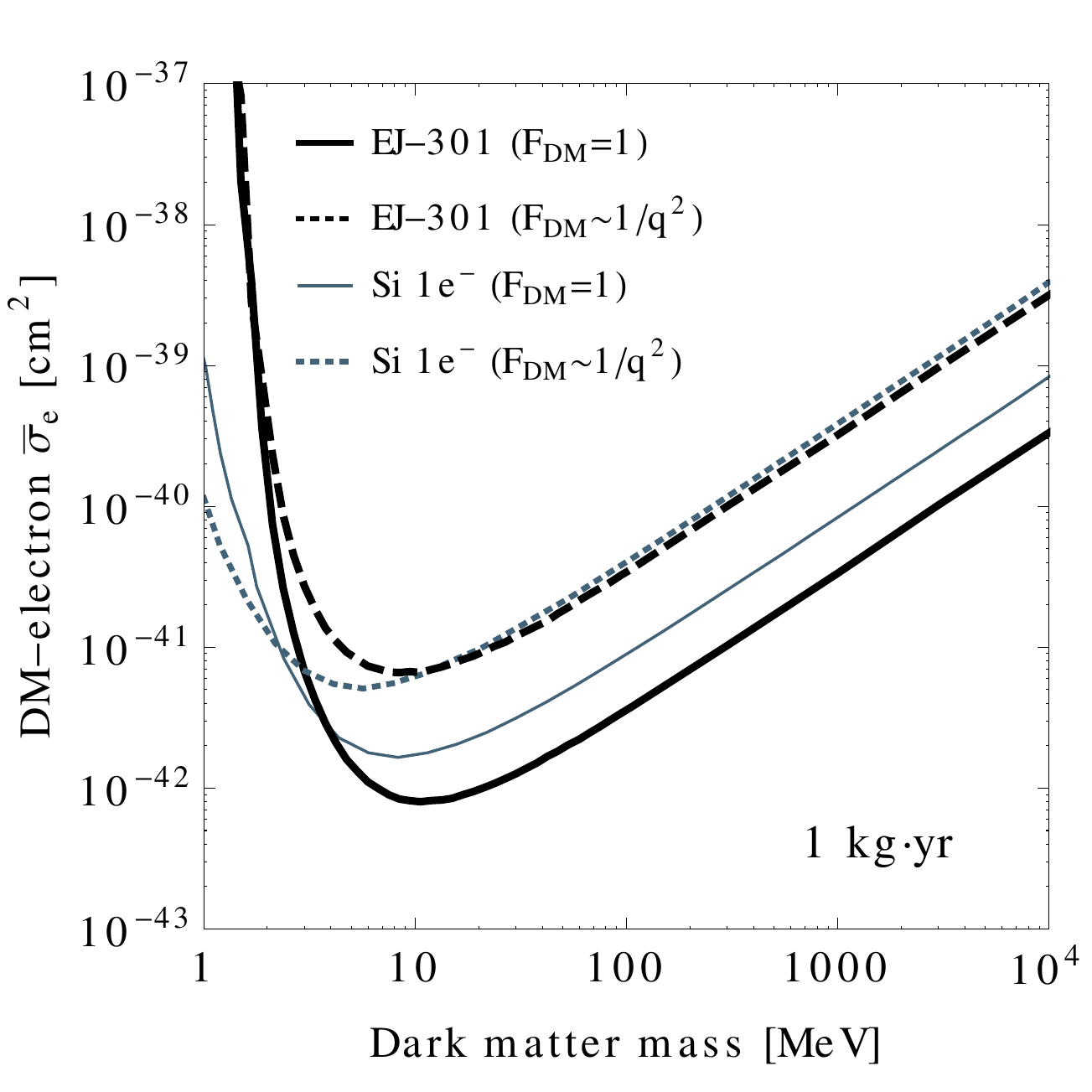}
    \caption{Comparison of EJ-301 and silicon reach with equal exposures. The 95\% C.L. exclusion reach for a hypothetical detector with 1~$\text{kg}-\text{yr}$ exposure and zero background events for both materials is shown in solid lines for $F_\text{DM}=1$ and dashed lines for $F_\text{DM}=\alpha^2 m_e^2 / q^2$. For the silicon $\overline{\sigma}_e$ we use the $1e^-$ example from~\cite{Essig:2015cda}. %With equal exposures, EJ-301 is comparable to silicon to within an order of magnitude for DM heavier than 10 MeV, with both DM form factors.
    With equal exposures, EJ-301 is comparable to silicon for $F_\text{DM} \sim 1/q^2$ for DM heavier than 10~\MeV, and for $F_\text{DM}=1$ performs somewhat better for $m_\chi \gtrsim 5~\MeV$.
    }  
    \label{fig:silicon}
\end{figure}

\section{Future Prospects With Organic Scintillators}
\label{sec:Future}
Organic scintillators with lower thresholds and anisotropic responses are clear steps forward for the next iteration of this technology. Polycyclic aromatic compounds such as polyacenes are promising target candidates since their electronic structures are well described by the formalism described herein~\cite{murrell1963theory}, and their properties as scintillators are well studied as is the case for naphthalene and anthracene~\cite{Birks1959a,bell1948use}. It should also be noted that as the excitation threshold decreases, one can expect the form factor to have the bulk of its support at lower momentum since the excitation energy and location of form factor peaks are both inversely propositional to the characteristic length between nodes in the wavefunctions, which increase for polyacenes with increasing numbers of rings. Both of these factors would increase sensitivity to lighter DM. Since the first transition in our idealized benzene model is suppressed by symmetry, one should expect this suppression to be lifted for larger and more complex (less symmetric) molecules, which would further lower the effective threshold and improve sensitivity to light DM. 

The variety of aromatic molecules such as polyacenes, trans-stilbene, substituted benzenes, and oligophenyl chromophores, which are known to be good scintillators, provides a wealth of options in selecting possible detector targets. By running more than one experiment with different targets it may also be possible to discriminate backgrounds inherent in the main scintillator target since the expected rate for a given DM mass and cross section will be different for each molecule. Since these organic compounds are relatively inexpensive and the target is distinct from the photodetector, using several targets is only a marginal extension to using a single target. 

Furthermore, the anisotropic nature of the crystal phases of these aromatic compounds can be used to look for directional modulation of the DM flux, further improving sensitivity to the DM--electron cross section and background rejection capabilities. In the solid state, large single crystals of polyacenes can be synthesized readily for pure compounds and binary/mixed scintillators~\cite{selvakumar2005growth,lipsett1957production,balamurugan2006scintillation}. These binary solid state targets maintain very high light yields while the crystal structure is known to produce anisotropic scintillation responses~\cite{tsukada1962directional,tsukada1965directional,brooks1974directional}. Aromatic liquid crystals based on anthracene core moieties, which can be aligned dynamically with electric fields while maintaining fluorescence quantum yield above 40\% and threshold around 2.5 eV, can also be made in the laboratory~\cite{gimenez2004luminescent}. Using this technology, one could imagine constructing a detector which tracks the DM wind by a continuous modulation of the electric field rather than a physical rotation of a crystal.

\section{Conclusions}
\label{sec:Conclusions}

In this paper, we have shown that organic scintillators are appealing targets for DM--electron scattering. The impressive reach, even for a non-optimized experiment, is due to two main factors. First, kilogram for kilogram, light DM scattering in aromatic hydrocarbons is as efficient as scattering in silicon for DM heavier than 10 MeV, even exceeding its sensitivity in the case of the $F_\text{DM}=1$ form factor. We illustrate this in Fig.~\ref{fig:silicon}, which compares the theoretical reach for 1 kg-yr of each material assuming a 100\% quantum efficiency for signal detection (we still take a 77\% radiative efficiency for EJ-301). We note that we cannot directly compare the form factors for p-xylene and silicon because the former consists only of discrete transitions, while the latter has a continuous band structure, so instead we integrate over the form factor and the DM velocity distribution, which is effectively a calculation of the rate per unit mass.\footnote{We note again that including electron ionization and associated secondary scintillation would increase the EJ-301 rate further, bring it closer to silicon.} Second, the scintillation process separates the primary scattering event (electron excitation) from the detected signal (a propagating scintillation photon), allowing for a separation of the target material from the detector. The scintillation photon, with energy of 2.9 eV, is in the near UV and can easily be transmitted across an interface (for example, an optical window) from the scintillator cell to any chosen photodetector. Other signals, such as electron/hole pairs and phonons, involve considerably more engineering at the interface in order to transmit the signal efficiently. The only requirement is that the area of the photodetector be comparable to the area of the optical window in the scintillator cell, for maximum light collection efficiency.

These two factors immediately suggest a straightforward way to improve the reach of a liquid scintillator search.\footnote{We are grateful to Noah Kurinsky for suggesting the improvements we propose here.} By reading out the scintillation signal with a low-background silicon photodetector, either the phonon-based sensor used in CDMS-HVeV or the skipper CCD used in SENSEI, we can take advantage of the best aspects of both setups: the large target mass (and low cost) of liquid scintillator, coupled to the low dark rate of the photosensor. Indeed, the reason our limits are competitive with the small-scale silicon experiments is that the effective background rate per kg of target material is very similar; the two-electron rate in SENSEI of $4.27 \times 10^{-5}$/pixel/day with an active mass of 0.0947 g is equivalent to 4.7 Hz/kg, which is comparable to the 13.6 Hz/kg intrinsic background rate per unit mass of the EJ-301 (i.e. before accounting for the 21.5\% efficiency of the PMT). The SENSEI sensitivity can obviously be improved by increasing the active mass, but assuming our background is not intrinsic to the scintillator itself, we could achieve similar limits by coupling the existing SENSEI detector to a kg scintillator cell, without the necessity of scaling up to a kg-scale mass of CCDs.\footnote{A potential concern with using silicon CCDs is that, since every scintillation photon will generate a single charge in a CCD pixel, high-energy background events producing many scintillation photons would appear as simultaneous single-electron events in multiple CCD pixels. This could be mistaken for a large single-electron rate, if the CCD charge integration time is too long. However, this can be avoided with a sufficiently high CCD readout rate. In SENSEI, the continuous readout integration time is 20 ms for 800 samples with an RMS noise of 0.14 electrons \cite{Crisler:2018gci}.  In shielded, low-background conditions, a majority of CCD frames would not contain signals from high-energy backgrounds in the scintillator, avoiding this concern at the expense of a modest dead time.} Further improvements would be possible by multiplexing several scintillator cells to achieve 10 kg of target mass, which, with zero background, would already be sensitive to a variety of thermal and non-thermal production mechanisms for sub-GeV DM \cite{Battaglieri:2017aum}. In future work, we plan to measure the intrinsic background of low-background EJ-301 (processed with ion-exchange resins for uranium and thorium removal) with silicon photodetectors in the NEXUS facility in the MINOS cavern at Fermilab \cite{NEXUS}, which has 300 m.w.e. overburden and a 15 mK dilution refrigerator which allows the readout stage to be at cryogenic temperatures to reduce the dark rate.

In conclusion, we believe that aromatic organic scintillators are a promising addition to the panoply of novel condensed matter systems suitable for DM--electron scattering. By leveraging the advances in low-background photodetectors and reducing the intrinsic background in the scintillator as much as possible, we propose that a large-exposure experiment with directional detection capabilities sensitive to well-motivated DM parameter space can be achieved with organic scintillators coupled to large-area, rather than large-mass, photodetectors.

\textbf{Acknowledgments.} We thank Daniel Baxter, Rouven Essig, Noah Kurinsky, Oren Slone, and Stuart Raby for helpful conversations, and C. Hurlbut for providing a EJ-301 sample for chemical analysis. We especially thank the organizers of the workshop ``New Directions in the Search for Light Dark Matter Particles'' and the Gordon and Betty Moore Foundation for providing a stimulating atmosphere for discussion. This work was supported in part by the Kavli Institute for Cosmological Physics at the University of Chicago through an endowment from the Kavli Foundation and its founder Fred Kavli. CB is supported by the US National Science Foundation Graduate Research Fellowship under grants number DGE-1144082 and DGE-1746045. Detector development was originally supported by NSF grant PHY-1506357.

\appendix

\section{Wavefunction and radiative efficiency calculations}
\label{app}

\subsection{Benzene wavefunction coefficients}
\label{sec:coeff}
When the core hamiltonian, $\mathcal{H}_{core}$, is diagonalized, the LCAO coefficients in Eq. (3) are given by the following~\cite{murrell1963theory},
\begin{equation}
\begin{aligned}
c_{2}^{j}  &=\frac{(1, 1, 1, 1, 1, 1 ) }{\sqrt{6}}  
&
c_{-2}^{j}  &=\frac{(-1, 1, -1, 1, -1, 1) }{\sqrt{6}} 
\\ 
c_{1}^{j}  &=\frac{(1, 0,-1,-1, 0,1 ) }{2} 
&
c_{-1^{\prime}}^{j} &=\frac{(-1, 2, -1, -1, 2, -1)}{\sqrt{12}}  \\ 
c_{-1}^{j}  &=\frac{(-1,0, 1, -1, 0, 1)}{2}  
&
c_{(1^{\prime})}^{j}  &=\frac{ (1, 2, 1, -1, -2, -1)}{\sqrt{12}}, 
\end{aligned}
\end{equation}
 where each $c^j$ is a collection of coefficients which gives an LCAO eigenstate. The normalization of these coefficients is such that each HMO is normalized to unity. 

When these HMOs are transformed into momentum space, the phase prefactors in Eq. (14) are given by the following:
\begin{equation}
    \begin{aligned}
    &\mathcal{B}_{2}= \sqrt{\frac{2}{3}} \left\{2 \cos \left(\frac{\sqrt{3}\;a k_{x} }{2}  \right) \cos \left(\frac{a k_{y}}{2}\right)+\cos (a k_{y})\right\} \\ 
    &\mathcal{B}_{1}=-2 i \left\{ \sin \left(\frac{\sqrt{3}\;a k_{x} }{2}\right) \cos \left(\frac{a k_{y}}{2}\right) \right\} \\
    &\mathcal{B}_{1\prime} = - i \frac{2}{\sqrt{3}} \left\{\cos \left(\frac{\sqrt{3}\;a k_{x} }{2}\right) \sin \left(\frac{a k_{y}}{2}\right)+\sin (a k_{y})\right\}\\
    &\mathcal{B}_{-1}=2 \left\{ \sin \left(\frac{\sqrt{3}\;a k_{x} }{2}\right) \sin \left(\frac{a k_{y}}{2}\right) \right\} \\
    &\mathcal{B}_{-1\prime} = \frac{2 }{\sqrt{3}} \left\{\cos (a k_{y})-\cos \left(\frac{\sqrt{3}\;a k_{x} }{2}\right) \cos \left(\frac{a k_{y}}{2}\right)\right\}\\
    &\mathcal{B}_{-2}= i \sqrt{\frac{2}{3}} \left\{2 \cos \left(\frac{\sqrt{3}\;a k_{x} }{2}\right) \sin \left(\frac{a k_{y}}{2}\right)-\sin (a k_{y})\right\},
    \end{aligned}
\end{equation}
where $a=0.14$ nm is the bond length of the aromatic carbon--carbon bond in benzene and p-xylene.

\subsection{Second and Third Singlet Excitations}
\label{sec:excit}

The second singlet excitations, in which a single electron is excited across the second lowest energy gap, are given by the following:
\begin{equation}
\begin{aligned}
    \psi_{1}^{s_2} &= 1/\sqrt{2}\; (\psi_{-1}^{2}+\psi_{-2}^{1}),\; \Delta E_{1}^{s_2} = 8.18~\eV\\
    \psi_{2}^{s_2} &= 1/\sqrt{2}\; (\psi_{-2}^{1\prime}+\psi_{-1\prime}^{2}),\; \Delta E_{1}^{s_2} = 8.18~\eV\\
    \psi_{3}^{s_2} &= 1/\sqrt{2}\; (\psi_{-2}^{1\prime}-\psi_{-1\prime}^{2}),\; \Delta E_{2}^{s_2} = 8.89~\eV\\
    \psi_{4}^{s_2} &= 1/\sqrt{2}\; (\psi_{-1}^{2}-\psi_{-2}^{1}),\; \Delta E_{2}^{s_2} = 8.89~\eV.
\end{aligned}
\label{eq:secondexcited}
\end{equation}
Notice that the mixing of configurational excited states results in an energy splitting around the quadruple degenerate second energy gap. It is found that these are the only linear combinations that give such a splitting. 

Finally, the third singlet excitation is given by 
\begin{equation}
\psi^{s_3} = \psi_{-2}^{2},\; \Delta E^{s_3} = 9.8~\eV.
\label{eq:thirdexcited}
\end{equation}

\subsection{Form factor details}
\label{sec:ffdetails}
%  \yk{Ben's pretty plots of form factors}

In this section we discuss the momentum dependence of the molecular form factors, including their angular profiles.
Although the angular dependence of the scattering rate is washed out for a detector in the liquid phase, a crystalline scintillator would provide an opportunity to use directionality to discriminate a dark matter signal from the background.

The nine lowest-lying transitions described in Eqs.~(\ref{eq:firstexcited}), (\ref{eq:secondexcited})~and~(\ref{eq:thirdexcited}) share a common feature: their form factors vanish when the imparted momentum $\vec{q}$ is orthogonal to the plane of the benzene ring.
%, and tend towards maxima when $\vec{q}$ lies in the plane.
Adjusting the orientation of the molecule with respect to the dark matter wind can thus have a strong effect on the total scattering rate.

For concreteness, we use a coordinate system where the six carbon nuclei in the benzene ring are located in the $z=0$ plane at:
\begin{align}
    \vec{R}_j = a \cos\left( j \frac{\pi}{3} - \frac{\pi}{6} \right) \vec{\hat{x}} +  a \sin\left( j \frac{\pi}{3} - \frac{\pi}{6} \right) \vec{\hat{y}} 
\end{align}
for $j=1,2,\ldots 6$,
where $a=0.14 \ \text{nm}$ is the bond length of the aromatic ring in both benzene and p-xylene.
All nine form factors exhibit the $\phi \rightarrow \phi + \frac{n \pi}{3}$ and $\phi \rightarrow -\phi$ symmetries of the benzene ring.
%We use $\theta_q$ and $\phi_q$ to describe the polar and azimuthal orientation of $\vec{q}$, respectively.

\begin{figure}[t]
    \centering
        \includegraphics[clip, trim=2cm 0cm 1cm 0cm,width=88mm]{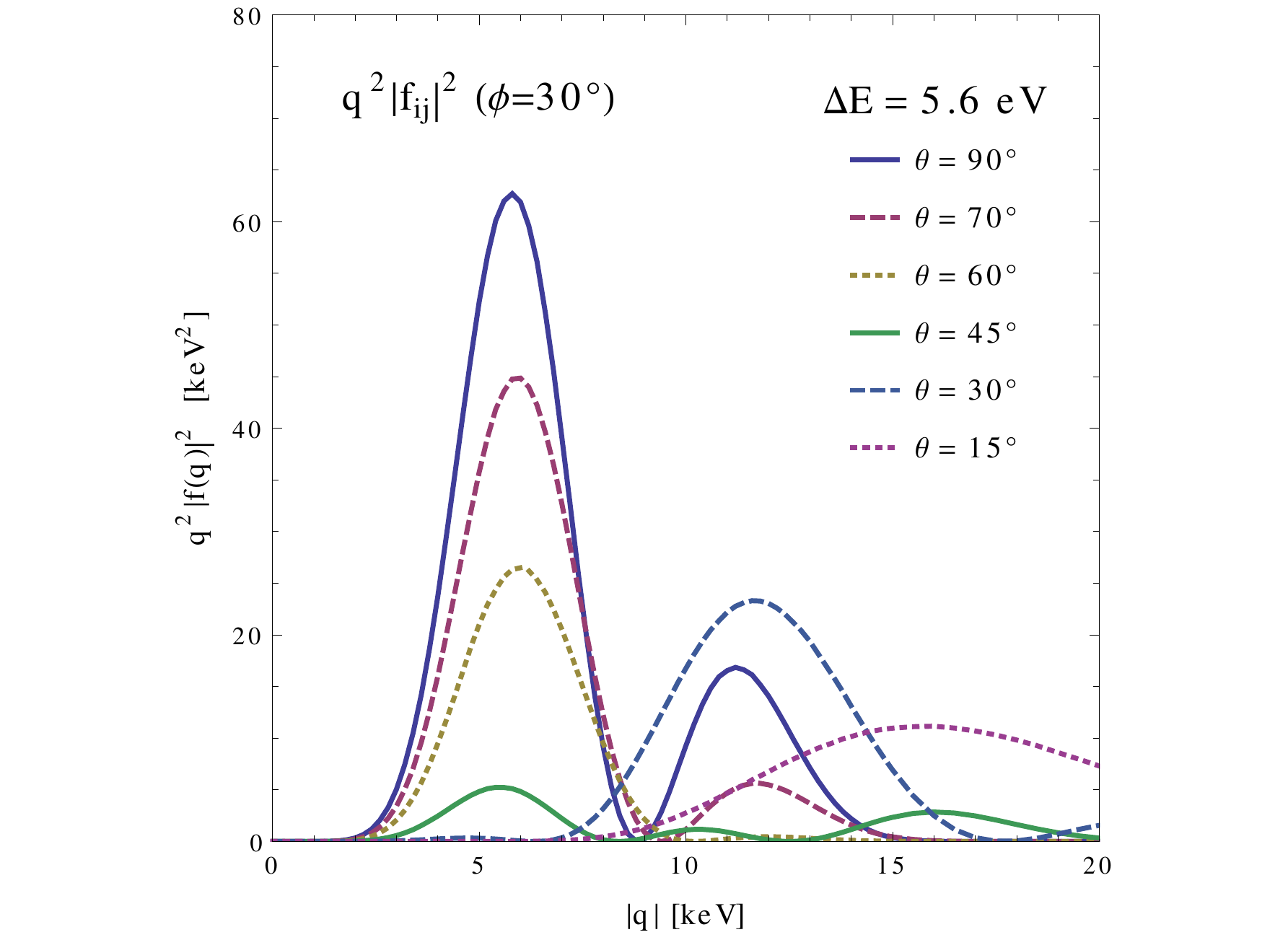}
        % \hspace{-20mm}
    \includegraphics[clip, trim=2cm 0cm 1cm 0cm,width=88mm]{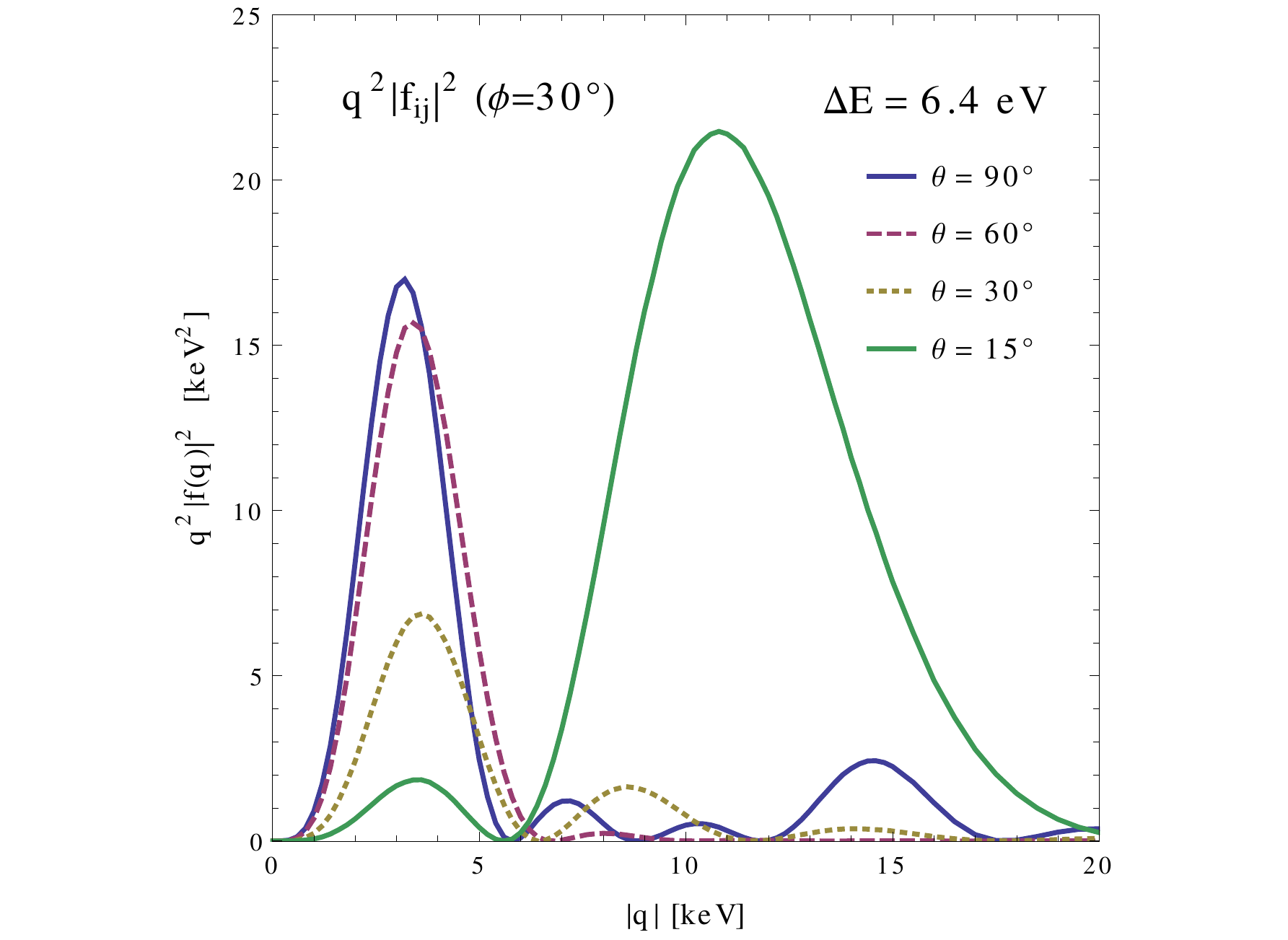}
    \caption{Angular profiles of the $5.6~\eV$ (top) and $6.4~\eV$ (bottom) form factors $q^2 \abs{f_{ij}(\vec{q})}^2$ for various values of the polar angle $\theta_q$. The azimuthal angle is fixed at $\phi_q = 30^\circ$, which in our coordinate system is parallel to the edges of the hexagonal ring. Both form factors reach maxima at $\theta = \frac{\pi}{2}$ where $q_z = 0$, and vanish in the $\theta \rightarrow 0$ ($\vec{q} \rightarrow q_z \vec{\hat{z}}$) limit. The $6.4~\eV$ transition has an additional peak at higher momenta when $10^\circ \lesssim\theta_q \lesssim 20^\circ$. 
    Neglecting the perturbation from the methyl groups in para-xylene, the form factors inherit the $\phi \rightarrow \phi + \frac{\pi}{3}$ shift symmetry of the benzene ring.}
    \label{fig:azimuthslice}
\end{figure}

For $m_\chi < 20~\MeV$ the scattering rate is driven almost exclusively by the 5.6~\eV\ and 6.4~\eV\ transitions, and
in Fig.~\ref{fig:azimuthslice} we show how their respective molecular form factors vary with the polar angle $\theta_q$ for fixed azimuthal angle $\phi_q = \frac{\pi}{6}$. The $5.6~\eV$ form factor is maximized at $\theta_q=\frac{\pi}{2}$ and $\phi_q = n\frac{\pi}{6}$ for integer $n$, where the momentum transfer $\vec{q}$ is parallel to the displacement between neighboring carbon nuclei: $\vec{q} \propto \vec{R}_i - \vec{R}_{i\pm1}$.

While the 5.6~\eV\ transition is driven mostly by in-plane scattering with $q \approx 6~\keV$, the 6.4~\eV\ transition is weighted towards lower momenta, peaked around $q \approx 3~\keV$ for most values of $\theta_q$. The 6.4~\eV\ transition is also responsive to a broad range of larger momenta $8~\keV \lesssim q \lesssim 15~\keV$, but only if the polar angle is relatively steep, with $\theta_q \approx 13^\circ$ as shown in the second panel of Fig.~\ref{fig:azimuthslice}.
Although the 5.6~\eV\ transition does exhibit some higher-momentum response at this $\theta_q \approx 15^\circ$ angle, for a broad range of momenta centered around $q \approx 16~\keV$, it affects the scattering rate to a lesser degree.

\begin{figure}[t]
    \centering
    \includegraphics[clip, trim=2cm 0cm 4cm 0cm,width=88mm]{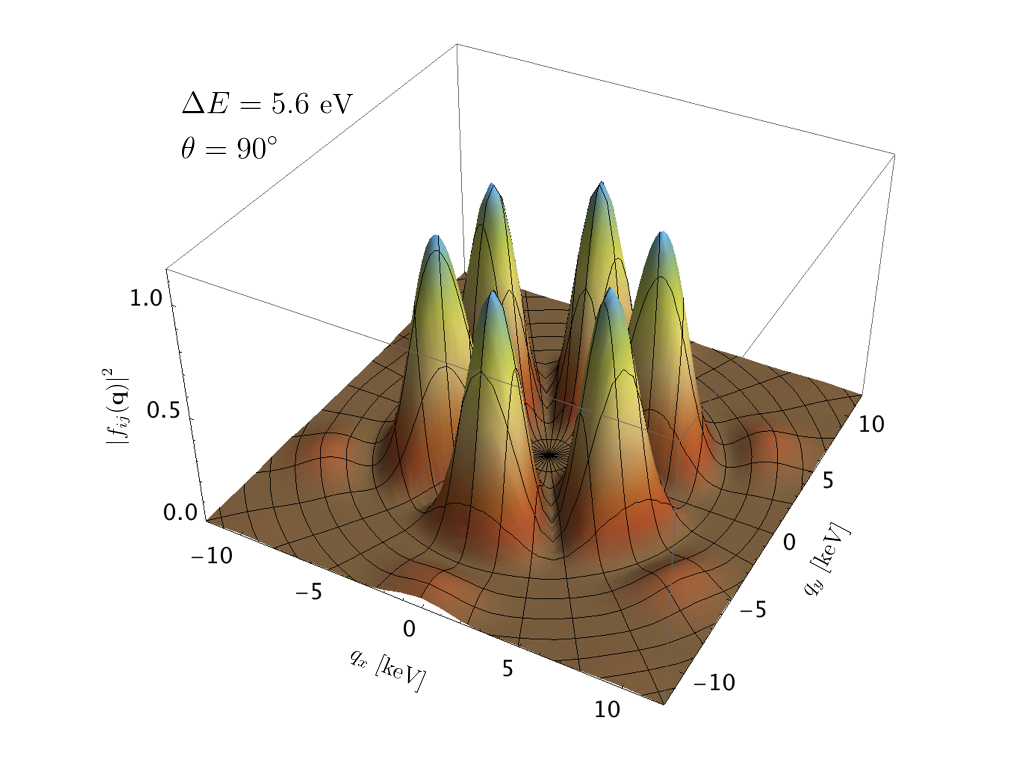}
        % \hspace{0mm}
    \includegraphics[clip, trim=2cm 0cm 4cm 0cm,width=88mm]{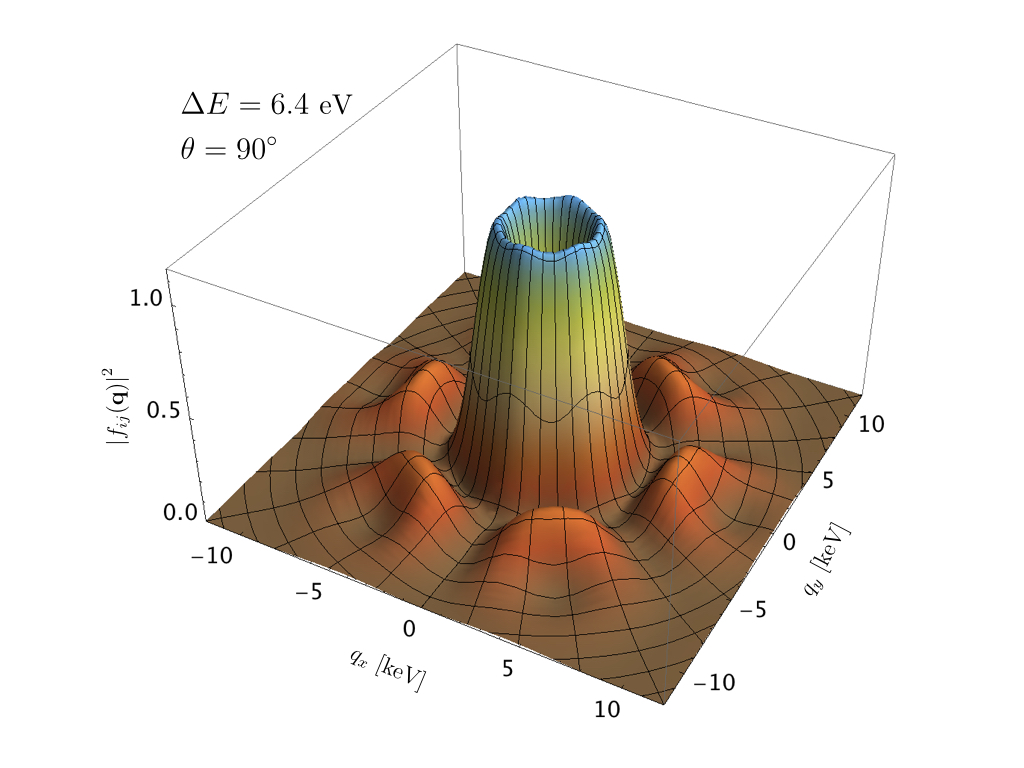}
    \caption{Azimuthal profiles of the $5.6~\eV$ (top) and $6.4~\eV$ (bottom) form factors $\abs{f_{ij}(\vec{q})}^2$, for momenta in the $q_z = 0$ plane. While the $5.6~\eV$ transition shows strong directionality in $\phi$, vanishing at $\phi = n\cdot 60^\circ$ for integer $n$, the sum of the two $6.4~\eV$ transitions is approximately rotationally invariant.}
    \label{fig:3dplots}
\end{figure}

Individual form factors also show strong directionality in the $\phi_q$ direction, but often in complementary ways that tend to average out. For example, adding together the $\abs{f_{ij}(\vec{q})}^2$ form factors for the two degenerate single-electron excitations that contribute to the $6.4~\eV$ transition, we find a nearly rotationally invariant form factor. However, the single $5.6~\eV$ transition retains its directionality, vanishing for $\phi_q = n \frac{\pi}{3}$ for integer $n$, and reaching sharp maxima at $\phi_q = \frac{\pi}{6} + n \frac{\pi}{3}$. %This behavior persists for $\theta$ above and below the plane.
Fig.~\ref{fig:3dplots} shows these two form factors along a particular conical slice at $\theta = \frac{\pi}{2}$, where the momentum $\vec{q}$ lies in the plane of the benzene ring. 

As demonstrated in Fig.~\ref{fig:molecularformfactor}, the presence of the $4.5~\eV$ transition is essentially irrelevant for DM--electron scattering, due to the strong suppression of its form factor. For completeness, it is worth commenting that its angular profile is complementary to that of the 5.6~\eV\ transition; that is, for integer $n$ its maxima occur at $\phi_q = n \frac{\pi}{3}$, and its form factor vanishes along $\phi_q = \frac{\pi}{6} + n \frac{\pi}{3}$ for all values of $\theta_q$.

% \begin{widetext}
% \end{widetext}

% \begin{widetext}

% \end{widetext}

If the dark matter mass $m_\chi$ is large enough to probe the higher momentum behavior of the form factor, then the presence of the secondary peak around $\theta_q \approx 13^\circ$ for momenta $q \gtrsim 8 \ \keV$ has potentially beneficial implications for directional detection of dark matter. In addition to the primary response around $\theta \approx 90^\circ$, an observation of the relatively sharp secondary peak at $\theta \approx 13^\circ$ could serve to confirm that a modulating signal originates from the scintillator and not some other source.
Furthermore, the location of the secondary peak in $\theta_q $ is a signature of benzene and its derivatives:  scintillators based on compounds with fundamentally different structures will generally have their own unique angular profile.

\subsection{Radiative efficiency calculation}
\label{sec:eff}

The radiative quantum yield of the scintillator is given by the following expression~\cite{1964a},
\begin{align}
Q       &=S \frac{ \omega }{P C \epsilon_{em}}\nonumber \\
        &= \left(\frac{Y \epsilon_{em}}{E_{abs}}\right) \frac{ \omega }{P C \epsilon_{em} } \\
        &= \frac{Y \omega }{P C E_{abs}}
\end{align}
where $S$ is the total scintillation efficiency, $P$ is the primary excitation efficiency, C is the energy efficiency (taken to be $2/3$ for organic scintillators), $\epsilon_{em}$ is the energy of the emitted scintillation photon, $\omega$ is the energy gap of the excitation, $Y$ is the experimentally measured light yield of the scintillator, and $E_\text{abs}$ is the total absorbed energy from an incoming particle (standardized to 1 MeV by light yield measurements). EJ-301 is measured by the manufacturer to have a light yield, $Y_{301}$, which is 78\% that of anthracene, $Y_\text{anth} = 1.74\times 10^4\; \text{photons}/\text{MeV}$~\cite{1964a}, and an excitation energy, $\omega_{301}$ of 4.5~eV. 
Note that using the lowest excitation in EJ-301 keeps our estimate conservative. Anthracene has an excitation energy of $\omega_\text{anth} = 3.15~\eV$~\cite{Birks1959a}. The excitation efficiency of organic molecules with aromatic rings is, $P\approx \frac{2}{3} F_{\pi}$, where $F_{\pi}$ is the fraction of $\pi$-electron in the molecule~\cite{1964a}. Here, we adopt values of the excitation efficiencies for EJ-301 and anthracene of $P_{301}$=0.098 and $P_\text{anth}$=0.01, respectively~\cite{1964a,daub1961proceedings}. We can now calculate the radiative quantum yield of EJ-301, $Q_{301}$ as a function of the documented radiative quantum yield of anthracene, $Q_\text{anth} = 0.68$~\cite{1964a, Birks1959a},

\begin{align}
    Q_{301} &= \frac{P_\text{anth}\;\omega_{301}\;Y_{301}}{P_{301}\; \omega_\text{anth}\; Y_\text{anth}} Q_\text{anth}  %\nonumber \\
            = 0.77.
\end{align}

 \subsection{Molecular Form Factor: Analytic Calculation} \label{appx:analytic}

The task of evaluating \eqref{eq:rate} with the molecular orbitals defined in \eqref{eq:Psitilde} is simplified by the fact that the benzene ring lies in a plane, allowing the $dp_z$ integral in \eqref{eq:molformfactor} to be completed by contour integration.
In this section we list the result, as well as the outline for an analytic method capable of providing $\abs{f_{ij}(\vec{q})}$ for an arbitrary arrangement of conjugated $2p_z$ orbitals aligned with the same $\hat{z}$ axis.

As described in Section~\ref{sec:Rate}, the LCAO momentum space molecular orbitals can be factored into a common $\tilde{\phi}(\vec{k})$, which is the Fourier transform of the $2p_z$ atomic orbital, and an orbital-specific prefactor $\mathcal B_\psi (\vec{k})$, which depends on the geometry of the molecule and the coefficients $c_i^\psi$. In terms of these two ingredients, the molecular form factor $f_{ij}(\vec{q})$ is described by
\begin{align}
    f_{if}(\vec{q}) = \int dp_x dp_y \mathcal B_i(p_x, p_y) \mathcal B_f^\star(p_x + q_x, p_y+q_y) \times \mathcal I_z,
\end{align}
where the newly defined $\mathcal I_z$ contains the $p_z$ dependence: 
\begin{widetext}
\begin{align}
\mathcal I_z &= \int dp_z  \tilde\phi(\vec{p}) \tilde{\phi}^\star(\vec{p} + \vec{q} ) 
= \frac{2}{\pi^2} \int_{-\infty}^\infty  \frac{dp_z\,p_z (p_z + q_z) }{\left( p_z^2 + p_x^2 + p_y^2 + \frac{1}{4} \right)^3
\left((p_z+q_z)^2 + (p_x+q_x)^2 + (p_y+q_y)^2 + \frac{1}{4}  \right)^3},
\label{eq:Izdimless}
\end{align}
where we have absorbed the factors of $a_0$ and $Z_\text{eff}$ into $\vec{q} = (q_x Z_\text{eff} a_0^{-1}, q_y Z_\text{eff} a_0^{-1}, q_z Z_\text{eff} a_0^{-1})$ and likewise for $\vec{p}$.
Introducing 
\begin{align}
f = \sqrt{ p_x^2 + p_y^2 + \frac{1}{4} } &&
g = \sqrt{ (p_x+q_x)^2 + (p_y+q_y)^2 + \frac{1}{4}}
\end{align}
as a convenient way to describe the locations of the poles in the complex plane, %at $p_z = \pm i f$ and $p_z = -q_z \pm i g$.
contour integration in $dp_z$ produces
\begin{align}
\mathcal I_z =& \frac{ (f+g) \left[ (f+g)^4(f^2 + 5 fg + g^2) 
 -2 q_z^2 (f+g)^2 (2 f^2 + 19 fg + 2 g^2) 
- 5 q_z^4 (f^2 - fg + g^2) \right]
}{4\pi f^3 g^3 \left( q_z^2 + (f+g)^2 \right)^5 } .
\label{eq:mathcalIzdef}
\end{align}
\end{widetext}

This analytic result significantly reduces the effort needed for the numeric integration, which can now be done over a smaller-dimensional volume.
With additional work, the remaining $dp_x dp_y$ integrals can be expressed instead in terms of generalized hypergeometric functions, based on the series expansion of $\mathcal I_z$ in the azimuthal direction defined by the coordinate transformation $p_x = \rho \cos \phi$, $p_y = \rho \sin \phi$.

% Taking both the coefficients $c_i^\psi$ and the locations of the nuclei $\vec{R}_i$ to be generic (though still requiring that $\vec{R}_i \cdot \hat{z} = 0$ for all $\vec{R}_i$), the form factor \begin{align}
% f_{ij}(\vec{q}) =& \int dp_x dp_y \left( \sum_{m,n} c^{(i)}_m e^{-i \vec{p} \cdot \vec{R}_m } c^{(j)\star}_n e^{+i (\vec{p} + \vec{q}) \cdot \vec{R}_n } \right) \mathcal I_z
% \end{align} 
% simplifies upon the definition of
% \begin{align}
%     R_{mn} = \abs{\vec{R}_m - \vec{R}_n} 
%     &&
% \end{align}
For greater generality, for the remainder of this section we take the locations of the nuclei $\vec{R}_i$ and the coefficients $c^j_\psi$ in the LCAO molecular orbitals to be generic: we require only that every $\vec{R}_i$ lies in the $z=0$ plane.
As we did for $\vec{p}$, we use cylindrical coordinates $q_z$, $q_\sigma$ and $q_\phi$ for the momentum transfer $\vec{q}$, with $q_x = q_\sigma \cos q_\phi$ and $q_y = q_\sigma \sin q_\phi$.

Given the coefficients $Z_k$ of the series expansion
\begin{equation}
    \mathcal I_z = \sum_{k=0} \frac{Z_k}{k!} \cos^k\left( \phi - q_\phi\right),
    \label{eq:seriesIz}
\end{equation}
we proceed to derive a general expression for the molecular form factor in terms of the Kamp\'e~de~Feri\'et hypergeometric function.

% \begin{align}
% f_{ij}(\vec{q}) =& \int dp_x dp_y \left( \sum_{m,n} c^{(i)}_m e^{-i \vec{p} \cdot \vec{R}_m } c^{(j)\star}_n e^{+i (\vec{p} + \vec{q}) \cdot \vec{R}_n } \right) \mathcal I_z
% \end{align} 

With this coordinate choice, $f_{ij}(\vec{q})$ has the generic form
\begin{equation}
f_{ij} (\vec{q}) = \sum_{mn} c_m^{(i)} c_n^{(j)} e^{i \vec{q}\cdot\vec{R}_j } \int \rho d\rho \int_{-\pi}^{\pi}d\varphi\, e^{i \rho R_{mn} \sin \varphi } \mathcal I_z,
\label{eq:formfunone}
\end{equation}
where for each pair $(m,n)$ we define $R_{mn} = Z_\text{eff}^{-1} a_0 \abs{\vec{R}_m - \vec{R}_n}$ as the dimensionless distance between the $m$th and $n$th nuclei, taking the angular separation of the two nuclei with respect to the center of the ring to be $\phi_{mn}$. 
A benzene-like ring with uniform radius $\abs{\vec{R}}$ 
satisfies
\begin{equation}
    \abs{\vec{R}_m - \vec{R}_n} = R \sqrt{2 - 2\cos\phi_{mn} }.
\end{equation}
Lastly we have defined
\begin{equation}
    \varphi \equiv \phi - \phi_{mn} + \tfrac{\pi}{2}
\end{equation}
so that \eqref{eq:formfunone} reduces to the integral form of a Bessel function when $\mathcal I_z$ is replaced by its series expansion.
Defining
\begin{align}
\mathcal I_\phi =\frac{1}{2\pi}\int_{-\pi}^{\pi}d\varphi\, e^{i \rho R_{mn} \sin \varphi } \mathcal I_z,
\end{align}
the resulting expression for the $d\varphi$ integral can be simplified to
\begin{widetext}
\begin{align}
\mathcal I_\phi &= \sum_{\gamma = 0}^\infty \left[ 2 - \sinc(\gamma \pi ) \right]  \left( \frac{i}{2} \right)^\gamma  J_\gamma(x)  \cos(\gamma (\phi_{mn} - q_\phi ))
\sum_{k'=0}^\infty \frac{2^{ - 2k'} Z_{\gamma + 2 k' } }{(k')! (k' + \gamma)!} ,
\end{align}
after some manipulation of the indices in the series expansions. The appearance of the $\sinc(\gamma \pi)$ is simply to avoid double-counting the $\gamma = 0$ case.
\end{widetext}

The remaining $d\rho$ integration depends on the functional form of $Z_n$. Rather than tackling the most generic case, where $\tilde{\phi}(\vec{k})$ is an arbitrary momentum space atomic wavefunction expressed in terms of a Gegenbauer polynomial and spherical harmonics, 
we continue to specialize to the $2p_z$ orbital and we evaluate $Z_n$ from \eqref{eq:mathcalIzdef} using a Taylor series.

As $f = \sqrt{\rho^2 + 1/4}$ is constant with respect to $\cos(\phi - q_\phi)$, the $Z_n$ can be determined from derivatives with respect to $g =  \sqrt{f^2 + 2 \rho q_\sigma \cos(\phi - q_\phi)} $:
\begin{equation}
    Z_n = \left. (\rho q_\sigma)^n \left( \frac{1}{g} \frac{d}{dg} \right)^n \mathcal I_z \right|_{g\rightarrow f}.
\end{equation}
Furthermore, even for relatively large $n$ the coefficients $Z_n$ can be calculated recursively with relatively good efficiency by noting that $\mathcal I_z$ and its $n$th derivative follow the general form
\begin{equation}
    Z_n(g) \sim g^\ell \left(\frac{1}{q_z^2 + (f+g)^2 } \right)^{5 + \alpha}
\end{equation}
for integers $\alpha \geq 0$ and $\ell$. Each term in $Z_n$ has an integer-valued coefficient $\MM_{\ell \alpha}$, such that 
\begin{align}
Z_n &=  \sum_{k} \frac{ \rho^n }{4\pi}(q_\sigma)^n (q_z^2)^k \MM_{nk} f  \left(\frac{1}{f^2}\right)^{k} \left( \frac{1}{q_z^2 + 4 f^2 } \right)^{5 + n}.
\end{align}

To perform the $d\rho$ integral in \eqref{eq:formfunone}, it is sufficient to calculate integrals of the form:
\begin{widetext}
\begin{align}
\mathcal I_{\gamma\ell k} &=
\int_0^\infty \frac{\rho d\rho \, \rho^{\gamma + 2\ell} J_\gamma(\rho R_{mn} ) }{R_{mn}^{\gamma + 2\ell+   2k + 7} }
\left( \frac{1}{\rho^2 + \frac{1}{4} } \right)^{k -1/2} \left( \frac{4}{q_z^2 + 4 \rho^2 + 1 } \right)^{5 + \gamma + 2\ell},
\label{eq:gellk}
\end{align}
where the factor of $R_{mn}$ has been added to make $\mathcal I_{\gamma \ell k}$. In a moment we discuss our analytic result for $\mathcal I_{\gamma \ell k}$; first, we show how $\mathcal I_{\gamma \ell k}$ appears in the molecular form factor:
\begin{align}
f_{ij}(\vec{q}) =\sum_{mn} c_m^{(i)} c^{(j)}_n e^{i \vec{q} \cdot \vec{R_n} } \sum_{\gamma = 0}^{\infty} \left[ 1 - \frac{\sinc(\gamma \pi )}{2} \right] \left( \frac{i}{2} \right)^\gamma \cos(\gamma (\phi_{mn} - q_\phi )) \sum_{\ell,k = 0}^\infty 
\frac{(q_\sigma)^{\gamma + 2 \ell} (q_z^2)^k  \MM_{\gamma + 2\ell, k} }{ \ell! (\gamma + \ell)!}
\frac{(R_{mn})^{ \gamma + 2\ell + 2k + 7} \mathcal I_{\gamma \ell k} }{4^{\gamma + 3\ell+5} }.
\end{align}

The coefficients $\MM_{n,k}$ can be calculated directly from a tensor product. From the derivatives of $\mathcal I_z$, we define a
\begin{align}
\BB(\alpha)^{\ell j}_{mn} = \delta^\ell_m \delta^j_n (\ell - 13 - 2 \alpha) + \delta_m^{\ell-1} \delta^j_n (2 \ell - 16 - 2\alpha) + 
\delta_m^{\ell - 2} \delta^j_n(\ell - 3)
+ \delta_m^{\ell - 2} \delta^{j+1}_n(\ell - 3)
\end{align}
such that 
\begin{equation}
Z_n =  \left. \frac{(\rho q_\sigma)^n}{4\pi f^9 f^{2n}} \frac{f^3}{g^3} \left(  \frac{ f^2}{q_z^2 + (f+g)^2 } \right)^{5+n} \left\{ \left(\frac{g}{f}\right)^m \left[\BB^{(n)}\right]^{\ell j}_{m k}  \beta_{\ell j} \right\} \left(\frac{q_z^2}{f^2}\right)^k  \right|_{g \rightarrow f},
\end{equation}
where
$\BB^{(n)}$ indicates the product $\BB(n-1) \cdot \ldots \BB(1) \cdot \BB(0)$,
and
where $\beta_{\ell j}$ encodes the coefficients in the polynomial in the numerator of \eqref{eq:mathcalIzdef}:
\begin{equation}
\beta_{\ell j} \left( \frac{g}{f} \right)^\ell \left( \frac{q_z^2}{f^2} \right)^j =
    \frac{f+g}{f^7} \left[ (f+g)^4(f^2 + 5 fg + g^2) 
 -2 q_z^2 (f+g)^2 (2 f^2 + 19 fg + 2 g^2) 
- 5 q_z^4 (f^2 - fg + g^2) \right].
\end{equation}
Inserting these particular values for $\beta_{\ell j}$,
%and applying the tensor multiplication
 $\MM$ is simply the $g \rightarrow f$ limit of the tensor product
% $\MM = \BB^{(n)} \cdot \beta$, we find
\begin{align}
\MM_{n, k} &=\left. \left(\frac{g}{f} \right)^m \left[\BB(n-1) \right]^{\ell_n j_n}_{m k} 
\left[\BB(n-2) \right]^{\ell_{n-1} j_{n-1} }_{\ell_{n} j_{n} } \ldots \left[\BB(1) \right]_{\ell_{2} j_2 }^{\ell_1 j_1} \left[\BB(0) \right]^{\ell j}_{\ell_1 j_1}   \beta_{\ell j}
  \right|_{g \rightarrow f}.
\end{align}
\end{widetext}

Finally, we turn our attention to evaluating $\mathcal I_{\gamma \ell k}$. First we rearrange the integral of \eqref{eq:gellk} using a Feynman parameter:
\begin{align}
    \frac{1}{A^a B^b} = \int_0^1 dx \frac{x^{a-1} (1-x)^{b-1} }{[x A + (1-x) B]^{a+b} } \frac{\Gamma(a+b)}{\Gamma(a) \Gamma(b)} 
\end{align}
where we take
\begin{align}
A &=  R_{mn}^2 (\rho^2 + 1/4),  
        &   a &= k -1/2 \\
B &= R_{mn}^2 (\rho^2 + 1/4 + q_z^2/4), 
        &   b &= n + 5 .
\end{align}
% For temporary convenience we revert to $n \equiv \gamma + 2\ell$.
Defining the parameter $u = R_{mn} \rho$, \eqref{eq:gellk} can be rewritten as
\begin{align}
\mathcal I_{\gamma\ell k} = &\int_0^1 dx \frac{\Gamma(k + \gamma + 2\ell + 9/2)  } {\Gamma(k-1/2) \Gamma(\gamma + 2\ell + 5 ) }  (1-x)^{\gamma + 2\ell+4} \nonumber\\
& \times x^{k-3/2}
 \int_0^\infty du   \frac{u^{\gamma + 2\ell+1} J_\gamma(u) }{[ u^2 + \Delta^2  ]^{k + \gamma + 2\ell + 9/2} } ,
\end{align}
where
\begin{equation}
    \Delta^2(x) = \frac{R_{mn}^2 }{4} + (1-x) \frac{R_{mn}^2 q_z^2}{4}.
\end{equation}
By replacing $J_\gamma(u)$ with its equivalent hypergeometric function ${_0 F_1}(-u^2/4)$, the $du$ integral can be completed:
% \begin{widetext}
\begin{align}
 \int_0^\infty du &  \frac{u^{\gamma + 2 \ell + 1} J_\gamma(u) }{[ u^2 + \Delta^2  ]^{k + \gamma + 2\ell + 9/2} }
 = 
 \frac{\left( \Delta^2 \right)^{-k - \ell - 7/2} }{2^{\gamma + 1}} 
 \nonumber \\ &\times 
  \frac{\Gamma(\ell + \gamma + 1) \Gamma(k + \ell + 7/2) }{\Gamma( \gamma + 1) \Gamma( k + \gamma + 2\ell + 9/2) } \;
\nonumber\\ &\times
{_1 F_2} \left( \left. \begin{array}{ c c} \ell + \gamma + 1 & \\ \gamma + 1 & -k - \ell - \frac{5}{2} \end{array} \right| \frac{\Delta^2 }{4} \right) ,
\end{align}
% \end{widetext}
where $_1 F_2$ is the hypergeometric function defined by the series
\begin{equation}
_1 F_2 \left[ \begin{array}{cc} a_1 &  \\ b_1 & b_2  \end{array}  ; x \right]  = \sum_{n=0}^{\infty} \frac{(a_1)_n }{(b_1)_n (b_2)_n } \frac{x^n}{n!} 
\end{equation}
using the standard notation where $(c)_n$ is the rising factorial
\begin{align}
(c)_n = 1\cdot \prod_{j = 0}^{n - 1} (c + j) = \frac{\Gamma(c+n)}{\Gamma(c)} .
\end{align}

Factoring out the remaining $dx$ integral as 
\begin{align}
\mathcal I_{\gamma \ell k}&= \frac{ 2^{-\gamma - 2\ell - 2k -8} \Gamma(\ell + \gamma + 1) \Gamma(k + \ell + 7/2)  }{\Gamma(k - 1/2 ) \Gamma( \gamma + 2\ell + 5) \Gamma(\gamma + 1) } \mathcal I_x,
\end{align}
\begin{align}
\mathcal I_x &=
\int_0^1 dx (1-x)^{\gamma + 2\ell + 4} x^{k - 3/2} \left( \frac{\Delta^2}{4} \right)^{-k -\ell - 7/2}
\nonumber\\ &\times
{_1 F_2} \left( \left. \begin{array}{ c c} \ell + \gamma + 1 & \\ \gamma + 1 & -k - \ell - \frac{5}{2} \end{array} \right| \frac{\Delta^2 }{4} \right) ,
\end{align}
we evaluate $\mathcal I_x$ by expanding 
the $_1 F_2$ function and $\Delta$ in powers of $x$, defining an $x_0$ and $\omega$ such that
\begin{equation}
\frac{\Delta^2}{4} = x_0 (1 - \omega x).
\end{equation}
Each term in the resulting series expansion is an Euler-type integral, which can be evaluated to produce
% \begin{widetext}
\begin{align}
\mathcal I_x & = x_0^{ - k - \ell- \frac{7}{2} } \frac{\Gamma(k - \frac{1}{2} ) \Gamma(\gamma + 2\ell + 5) }{\Gamma( k + \gamma + 2 \ell + \frac{9}{2} ) }  
\sum_{m,j=0}^\infty \left( \frac{(\ell + \gamma + 1)_m}{(\gamma + 1)_m} \right.
\nonumber\\ & \times \left.
\frac{(k - \frac{1}{2} )_j }{ (k + \gamma + 2 \ell + \frac{9}{2}  )_j } \frac{1}{(-k -\ell - \frac{5}{2} )_{m - j} } \frac{x_0^m }{m!} \frac{\omega^j }{j!}  \right).
\label{eq:pleaseletthisbeover}
\end{align}
% \end{widetext}

To return this infinite series as a closed form expression, 
we invoke the Kamp\'e~de~Feri\'et function---a two-argument generalization of the generalized hypergeometric function, which following the notation of~\cite{exton_1978}
has the series expansion
\begin{align}
F^{A: B; B'}_{C: D; D'} \left( \left.\left.\left. \begin{array}{c} a_1 \ldots a_A \\ c_1 \ldots c_C \end{array} \right|
\begin{array}{c} b_1 \ldots b_B \\ d_1 \ldots d_D \end{array} \right|
\begin{array}{c} b'_1 \ldots b'_{B'} \\ d'_1 \ldots d'_{D'} \end{array} \right| x, y \right)
\nonumber\\
=\sum_{m,n = 0}^\infty \frac{(\vec{a})_{m+n} }{(\vec{c})_{m+n} } \frac{(\vec{b})_m }{(\vec{d})_m } \frac{(\vec{b'})_n }{(\vec{d'})_n } \frac{x^m y^n }{m! n!},
\label{eq:KdF}
\end{align}
using the shorthand notation $(\vec{e})_r = \prod_i^E (e_i)_r$.
To match \eqref{eq:pleaseletthisbeover} to the Kamp\'e~de~Feri\'et function requires that we reindex the sum over $m$ and $j$. We split the double series into three regions: $m\geq j$; $m\leq j$; and $m=j$, so that to remove the double-counted $m=j$ entries we take
\begin{equation}
\sum_{m,j = 0 }^\infty = \sum_{m \geq j} + \sum_{j \geq m } - \sum_{m=j}.
\end{equation}

\begin{widetext}
Our result for $\mathcal I_x$ is
\begin{align}
\mathcal I_x = \left( \frac{1}{x_0} \right)^{k + \ell + \frac{7}{2}  } \frac{\Gamma( k - \frac{1}{2} ) \Gamma( \gamma + 2\ell + 5 ) }{\Gamma(k + \gamma + 2\ell + \frac{9}{2} ) } \left( \mathcal I_{m\geq j} +  \mathcal I_{j\geq m} -  \mathcal I_{m= j} \right),
\end{align}
defining the three functions as
\begin{align}
\mathcal I_{m \geq j} &= F^{1: 1; 1}_{2: 1; 1} \left( \left.\left.\left. \begin{array}{c c} \ell+\gamma+1 & \\ \gamma+1 & 1 \end{array} \right| \begin{array}{c} 1 \\ -k-\ell-\frac{5}{2} \end{array} \right| \begin{array}{c} k-\frac{1}{2} \\ k+\gamma+ 2\ell+\frac{9}{2} \end{array} \right| x_0, x_0 \omega \right) \\
\mathcal I_{j \geq m} &= F^{1:1; 2}_{2:1; 0}  \left( \left.\left.\left. \begin{array}{c c} k-\frac{1}{2} & \\ k+\gamma+2\ell+\frac{9}{2} & 1 \end{array} \right| \begin{array}{c} \ell+\gamma+1 \\ \gamma+1 \end{array} \right| \begin{array}{c c} k+\ell+\frac{7}{2} & 1 \\ \text{---} & \end{array} \right| x_0 \omega, -\omega \right) \\
\mathcal I_{j=m} &= {_2 F_3} \left( \left. \begin{array}{c c c} \ell+\gamma+1 & k-\frac{1}{2} & \\ \gamma+1 & k+\gamma+2\ell + \frac{9}{2} & 1 \end{array} \right| x_0\omega \right).
\end{align}
% and
% \begin{align}
% \omega = \frac{1}{1 + Z_\text{eff}^2/q_z^2 }
% &&
% x_0 = \frac{R_{ij}^2 Z_\text{eff}^2}{16} \left(  \frac{1}{\omega} \right).
% \end{align}
In certain special cases such as $\ell =0$, or more generally when a pair of coefficients in the numerator and denominator match, these functions simplify to lower-order hypergeometric functions. 
\end{widetext}

\bibliography{OrganicScintillator.bib}

\end{document}